\documentclass[preprint]{aastex}

\usepackage{amsmath,amssymb}
\usepackage{amsfonts,latexsym}

\usepackage{graphicx}
\usepackage[bookmarks=true,colorlinks=true,citecolor=blue,linkcolor=blue,urlcolor=magenta]{hyperref}

\usepackage{xcolor}

\usepackage{soul}
\definecolor{lightblue}{rgb}{.90,.90,1}

\begin{document}

\title{Thermomagnetic Ettingshausen-Nernst effect in tachocline\\and axion mechanism of solar luminosity variations}

\author{V.D.~Rusov\footnote{Corresponding author e-mail: siiis@te.net.ua}, I.V.~Sharph, V.P.~Smolyar, M.E.~Beglaryan}
\affil{Department of Theoretical and Experimental Nuclear Physics,\\ Odessa National Polytechnic University, Odessa, Ukraine}


\author{M.V.~Eingorn}
\affil{CREST and NASA Research Centers, North Carolina Central University,\\ Durham, North Carolina, U.S.A.}

\begin{abstract}
It is shown that the hypothesis of the axion mechanism for solar luminosity variations
suggesting that the solar axion particles are born in the core of the Sun
and may be efficiently converted back into $\gamma$-quanta in the magnetic
field of the solar overshoot tachocline is physically relevant. As a result, it
is also shown that the intensity variations of the $\gamma$-quanta of axion
origin, induced by the magnetic field variations in the tachocline via the 
thermomagnetic Ettingshausen-Nernst effect, directly cause the Sun luminosity 
variations and eventually characterize the active and quiet states of the Sun.

Within the framework of this mechanism estimations of the strength of the axion
coupling to a photon ($g_{a \gamma} = 4.4 \cdot 10^{-11} GeV^{-1}$) and the
hadronic axion particle mass ($m_a \sim 3.2 \cdot 10^{-2} eV$) have been
obtained. It is also shown that the claimed axion parameters do not contradict
any known experimental and theoretical model-independent limitations.
\end{abstract}

\section{Introduction}

A hypothetical pseudoscalar particle called axion is predicted by the theory
related to solving the CP-invariance violation problem in QCD. The most
important parameter determining the axion properties is the energy scale $f_a$
of the so-called U(1) Peccei-Quinn symmetry violation. It determines both the
axion mass and the strength of its coupling to fermions and gauge bosons
including photons. However, in spite of the numerous direct experiments, they
have not been discovered so far. Meanwhile, these experiments together with the
astrophysical and cosmological limitations leave a rather narrow band for the
permissible parameters of invisible axion (e.g.
$10^{-6} eV \leqslant m_a \leqslant 10^{-2} eV$~\citep{ref01,ref02}), which is
also a well-motivated cold dark matter candidate in this mass region
\citep{ref01,ref02}.

A whole family of axion-like particles (ALP) with their own features may exist
along with axions having the similar Lagrangian structure relative to the
Peccei-Quinn axion, as well as their own distinctive features. It consists in
the fact that if they exist, the connection between their mass and their
constant of coupling to photons must be highly weakened, as opposed to the
axions. It should be also mentioned that the phenomenon of photon-ALP mixing in
the presence of the electromagnetic field not only leads to the classic
neutrino-like photon-ALP oscillations, but also causes the change in the
polarization state of the photons (the $a \gamma \gamma$ coupling acts like a
polarimeter \citep{ref03}) propagating in the strong enough magnetic fields. It
is generally assumed that there are light ALPs coupled only to two photons,
although the realistic models of ALPs with couplings both to photons and to
matter are not excluded \citep{ref04}. Anyway, they may be considered a
well-motivated cold dark matter candidate \citep{ref01,ref02} under certain
conditions, just like axions.

It is interesting to note that the photon-ALP mixing in magnetic fields of
different astrophysical objects including active galaxies, clusters of
galaxies, intergalactic space and the Milky Way, may be the cause of the
remarkable phenomena like dimming of stars luminosity (e.g. supernovae in the
extragalactic magnetic field \citep{ref06,ref07}) and ``light  shining through
a wall'' (e.g. light from very distant objects, travelling through the Universe
\citep{ref03,ref05}). In the former case the luminosity of an astrophysical
object is dimmed because some part of photons transforms into axions in the
object's magnetic field. In the latter case photons produced by the object are
initially converted into axions in the object's magnetic field, and then after
passing some distance (the width of the ``wall'') are converted back into
photons in another magnetic field (e.g. in the Milky Way), thus emulating the
process of effective free path growth for the photons in astrophysical medium
\citep{ref08,ref09}.

For the sake of simplicity let us hereinafter refer to all such particles as
axions if not stated otherwise.

In the present paper we consider the possible existence of the axion mechanism
of Sun luminosity variations\footnote{Let us point out that the axion mechanism of Sun
luminosity used for estimating the axion mass was described for the first time
in 1978
by \cite{ref10}.} based on the ``light shining through a wall'' effect. To be
more exact, we attempt to explain the axion mechanism of Sun luminosity variations by the
``light shining through a wall'', when the photons born mainly in the solar
core are at first converted into axions via the Primakoff effect \citep{ref11}
in its magnetic field, and then are converted back into photons after passing
the solar radiative zone and getting into the magnetic field of the overshoot
tachocline. We estimate this magnetic field within the framework of the 
Ettingshausen-Nernst effect. In addition to that we obtain the consistent 
estimates for the axion mass ($m_a$) and the axion coupling constant to photons
($g_{a \gamma}$), basing on this mechanism, and verify their values against the
axion model results and the known experiments including CAST, ADMX, RBF.

\section{Photon-axion conversion and the case of maximal mixing}

Let us give some implications and extractions from the photon-axion
oscillations theory which describes the process of the photon conversion into
an axion and back under the constant magnetic field $B$ of the length $L$. It
is easy to show \citep{ref05,Raffelt-Stodolsky1988,ref07,Hochmuth2007} that in
the case of the negligible photon absorption coefficient
($\Gamma _{\gamma} \to 0$) and axions decay rate ($\Gamma _{a} \to 0$) the
conversion probability is
\begin{equation}
P_{a \rightarrow \gamma} = \left( \Delta_{a \gamma}L \right)^2 \sin ^2 \left( \frac{ \Delta_{osc}L}{2} \right) \Big/ \left( \frac{ \Delta_{osc}L}{2}
\right)^2 \label{eq01}\, ,
\end{equation}
where the oscillation wavenumber $\Delta_{osc}$ is given by
\begin{equation}
\Delta_{osc}^2 = \left( \Delta_{pl} + \Delta_{Q,\perp} - \Delta_{a} \right)^2 + 4 \Delta_{a \gamma} ^2
\label{eq02}
\end{equation}
while the mixing parameter $\Delta _{a \gamma}$, the axion-mass parameter
$\Delta_{a}$, the refraction parameter $\Delta_{pl}$ and the QED dispersion
parameter $\Delta_{Q,\perp}$ may be represented by the following expressions:
\begin{equation}
\Delta _{a \gamma} = \frac{g_{a \gamma} B}{2} = 540 \left( \frac{g_{a \gamma}}{10^{-10} GeV^{-1}} \right) \left( \frac{B}{1 G} \right) ~~ pc^{-1}\, ,
\label{eq03}
\end{equation}
\begin{equation}
\Delta _{a} = \frac{m_a^2}{2 E_a} = 7.8 \cdot 10^{-11} \left( \frac{m_a}{10^{-7} eV} \right)^2 \left( \frac{10^{19} eV}{E_a} \right) ~~ pc^{-1}\, ,
\label{eq04}
\end{equation}
\begin{equation}
\Delta _{pl} = \frac{\omega ^2 _{pl}}{2 E_a} = 1.1 \cdot 10^{-6} \left( \frac{n_e}{10^{11} cm^{-3}} \right) \left( \frac{10^{19} eV}{E_a} \right) ~~ pc^{-1},
\label{eq05}
\end{equation}
\begin{equation}
\Delta _{Q,\perp} = \frac{m_{\gamma, \perp}^2}{2 E_a} .
\label{eq06}
\end{equation}

Here $g_{a \gamma}$ is the constant of axion coupling to photons; $B$ is the
transverse magnetic field; $m_a$ and $E_a$ are the axion mass and energy;
$\omega ^2 _{pl} = 4 \pi \alpha n_e / m_e$ is an effective photon mass in terms
of the plasma frequency if the process does not take place in vacuum, $n_e$ is
the electron density, $\alpha$ is the fine-structure constant, $m_e$ is the
electron mass; $m_{\gamma, \perp}^2$ is the effective mass square of the
transverse photon which arises due to interaction with the external magnetic
field.

The conversion probability (\ref{eq01}) is energy-independent, when
$2 \Delta _{a \gamma} \approx \Delta_{osc}$, i.e.
\begin{equation}
P_{a \rightarrow \gamma} \cong \sin^2 \left( \Delta _{a \gamma} L \right)\, ,
\label{eq07}
\end{equation}
or, whenever the oscillatory term in (\ref{eq01}) is small
($\Delta_{osc} L / 2 \to 0$), implying the limiting coherent behavior
\begin{equation}
P_{a \rightarrow \gamma} \cong \left( \frac{g_{a \gamma} B L}{2} \right)^2\, .
\label{eq08}
\end{equation}

It is worth noting that the oscillation length corresponding to (\ref{eq07})
reads
\begin{equation}
L_{osc} = \frac{\pi}{\Delta_{a \gamma}} = \frac{2 \pi}{g_{a \gamma} B} \cong 5.8 \cdot 10^{-3}
\left( \frac{10^{-10} GeV^{-1}}{g_{a \gamma}} \right)
\left( \frac{1G}{B} \right)  ~pc
\label{eq13}
\end{equation}
\noindent assuming a purely transverse field. In the case of the appropriate
size $L$ of the region a complete transition between photons and axions is
possible.

From now on we are going to be interested in the energy-independent case
(\ref{eq07}) or (\ref{eq08}) which plays the key role in determination of the parameters
for the axion mechanism of Sun luminosity variations hypothesis (the axion coupling
constant to photons $g_{a \gamma}$, the transverse magnetic field $B$ of length
$L$ and the axion mass $m_a$).

\section{Axion mechanism of Sun luminosity variations}

Our hypothesis is that the solar axions which are born in the solar core
\citep{ref01,ref02} through the known Primakoff effect \citep{ref11}, may be
converted back into $\gamma$-quanta in the magnetic field of the solar
tachocline (the base of the solar convective zone). The magnetic field
variations in the tachocline cause the converted $\gamma$-quanta intensity
variations in this case, which in their turn cause the variations of the Sun
luminosity known as the active and quiet Sun states. Let us consider this
phenomenon in more detail below.

As we noted above, the expression (\ref{eq01}) for the probability of the
axion-photon oscillations in the transversal magnetic field was obtained for
the media with the quasi-zero refraction, i.e. for the media with a negligible
photon absorption coefficient ($\Gamma_{\gamma} \to 0$). It means that in order
for the axion-photon oscillations to take place without any significant losses,
a medium with a very low or quasi-zero density is required, which would
suppress the processes of photon absorption almost entirely.

Surprisingly enough, it turns out that such ``transparent'' media can take
place, and not only in plasmas in general, but straight in the convective zone
of the Sun. Here we generally mean the so-called magnetic flux tubes, the
properties of which are examined below.

\subsection{Ideal photon channeling conditions inside the magnetic flux tubes}

\label{subsec-channeling}

\begin{figure*}
\begin{center}
\includegraphics[width=12cm]{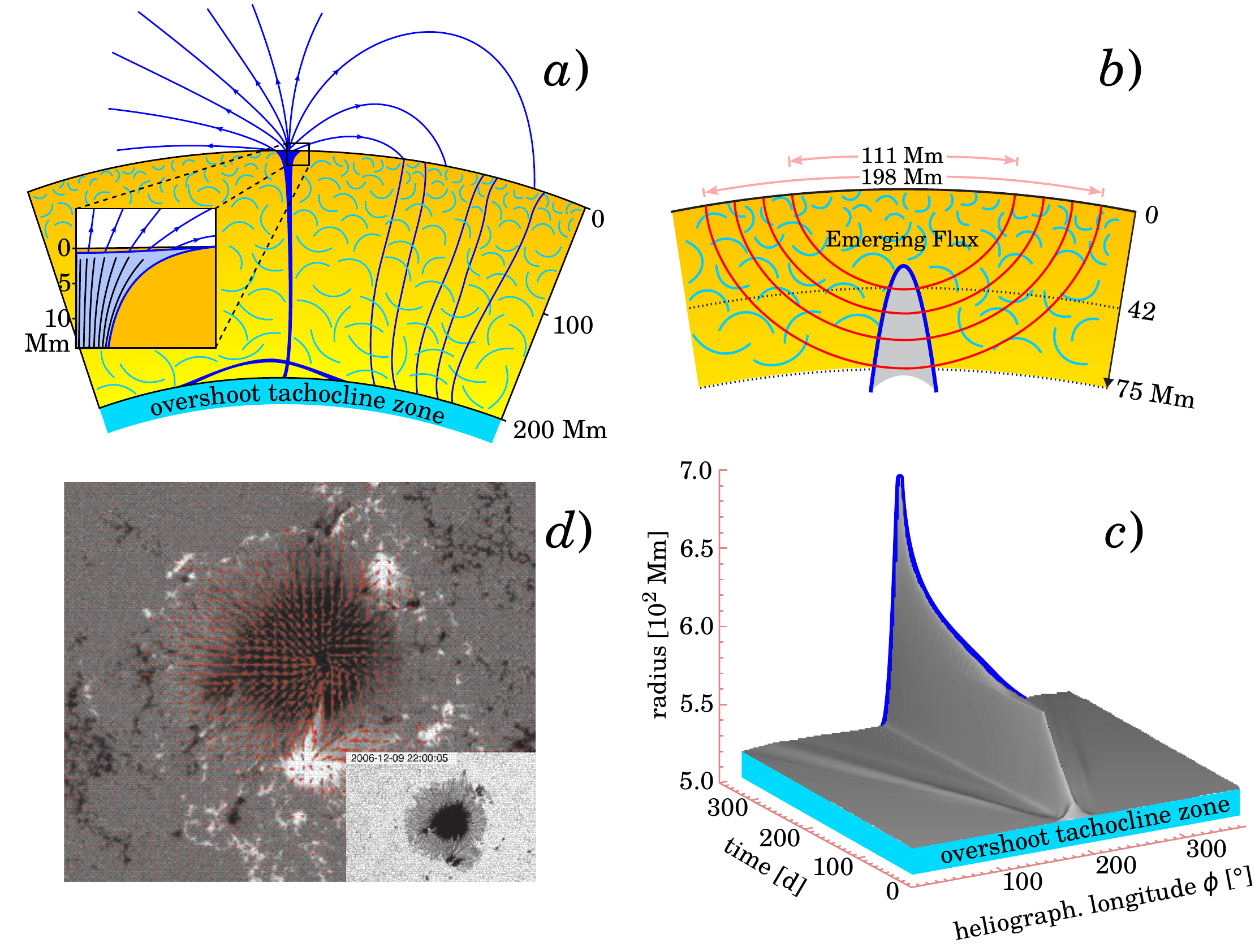}
\end{center}
\caption{(a) Vertical cut through an active region illustrating the connection
between a sunspot at the surface and its origins in the toroidal field layer at
the base of the convection zone. Horizontal fields stored at the base of the
convection zone (the overshoot tachocline zone) during the cycle. Active
regions form from sections brought up by buoyancy (one shown in the process of
rising). After eruption through the solar surface a nearly potential field is
set up in the atmosphere (broken lines), connecting to the base of the
convective zone via almost vertical flux tube. Hypothetical small scale
structure of a sunspot is shown in the inset (Adopted from
\cite{ref18}
and
\cite{ref15}).
(b) Detection of emerging sunspot regions in the solar interior~\citep{ref18}.
Acoustic ray paths with lower turning points between 42 and 75 Mm
(1 Mm=1000 km) crossing a region of emerging flux. For simplicity, only four
out of total of 31 ray paths used in this study (the time-distance
helioseismology experiment) are shown here. Adopted from~\cite{ref19}.
(c) Emerging and anchoring of stable flux tubes in the overshoot tachocline
zone, and its time-evolution in the convective zone. Adopted from \cite{ref20}.
(d) Vector magnetogram of the white light image of a sunspot (taken with SOT on
a board of the Hinode satellite -- see inset) showing in red the direction of
the magnetic field and its strength (length of the bar). The movie shows the
evolution in the photospheric fields that has led to an X class flare in the
lower part of the active region. Adopted from~\cite{ref21}.}
\label{fig01}
\end{figure*}

The idea of the energy flow channeling along a fanning magnetic field has been
suggested for the first time by
\cite{ref12} as an explanation for
darkness of umbra of sunspots. It was incorporated in a simple sunspot model by
\cite{ref13}.
\cite{ref14} extended this suggestion to smaller
flux tubes to explain the dark pores and the bright faculae as well.
Summarizing the research of the convective zone magnetic fields in the form of
the isolated flux tubes,
\cite{ref15} suggested a simple mathematical model for the behavior of thin
magnetic flux tubes, dealing with the nature of the solar cycle, sunspot
structure, the origin of spicules and the source of mechanical heating in the
solar atmosphere. In this model, the so-called thin tube approximation is used
(see \cite{ref15} and references therein), i.e. the field is conceived to exist
in the form of slender bundles of field lines (flux tubes) embedded in a
field-free fluid (Fig.~\ref{fig01}). Mechanical equilibrium between the tube
and its surrounding is ensured by a reduction of the gas pressure inside the
tube, which compensates the force exerted by the magnetic field.  In our
opinion, this is exactly the kind of mechanism
\cite{Parker1955} was thinking about when he wrote about the problem of flux
emergence: ``Once the field has been amplified by the dynamo, it needs to be
released into the convection zone by some mechanism, where it can be
transported to the surface by magnetic buoyancy''~\citep{ref17}.

In order to understand magnetic buoyancy, let us consider an isolated
horizontal flux tube in pressure equilibrium with its non-magnetic surroundings
so that 
in cgs units
\begin{equation}
p_{ext} = p_{int} + \frac{\vert \vec{B} \vert^2}{8 \pi} ,
\label{eq21}
\end{equation}
\noindent where $p_{int}$ and $p_{ext}$ are the internal and external gas
pressures respectively, $B$ denotes the uniform field strength in the flux
tube. If the internal and external temperatures are equal so that $T_{int} =
T_{ext}$ (thermal equilibrium), then since $p_{ext} > p_{int}$, the gas in the
tube is less dense than its surrounding ($\rho _{ext} > \rho _{int}$), implying
that the tube will rise under the influence of gravity.

In spite of the obvious, though turned out to be surmountable, difficulties of
the application to real problems, it was shown (see~\cite{ref15} and Refs.
therein) that strong buoyancy forces act in magnetic flux tubes of the required
field strength ($10^4 - 10^5 ~G$~\citep{ref23}). Under their influence tubes
either float to the surface as a whole (e.g. Fig.~1 in \citep{ref24}) or they
form loops of which the tops break through the surface (e.g. Fig.~1
in~\citep{ref14}) and lower parts descend to the bottom of the convective zone,
i.e. to the overshoot tachocline zone. The convective zone, being unstable,
enhanced this process~\citep{ref25,ref26}. Small tubes take longer to erupt
through the surface because they feel stronger drag forces. It is interesting
to note here that the phenomenon of the drag force which raises the magnetic
flux tubes to the convective surface with the speeds about 0.3-0.6~km/s, was
discovered in direct experiments using the method of time-distance
helioseismology~\citep{ref19}. Detailed calculations of the
process~\citep{ref27} show that even a tube with the size of a very small spot,
if located within the convective zone, will erupt in less than two years. Yet,
according to~\cite{ref27}, horizontal fields are needed in overshoot tachocline
zone, which survive for about 11~yr, in order to produce an activity cycle.

A simplified scenario of magnetic flux tubes (MFT) birth and space-time
evolution (Fig.~\ref{fig01}a) may be presented as follows. MFT is born in the
overshoot tachocline zone (Fig.~\ref{fig01}c) and rises up to the convective
zone surface (Fig.~\ref{fig01}b) without separation from the tachocline (the
anchoring effect), where it forms the sunspot (Fig.~\ref{fig01}d) or other
kinds of active solar regions when intersecting the photosphere.  There are
more fine details of MFT physics expounded in overviews by
\cite{ref17} and
\cite{ref24}, where certain fundamental questions, which need to be addressed
to understand the basic nature of magnetic activity, are discussed in detail:
How is the magnetic field generated, maintained and dispersed? What are its
properties such as structure, strength, geometry? What are the dynamical
processes associated with magnetic fields? \textbf{What role do magnetic fields
play in energy transport?}

Dwelling on the last extremely important question associated with the energy
transport, let us note that it is known that thin magnetic flux tubes can
support longitudinal (also called sausage), transverse (also called kink),
torsional (also called torsional Alfv\'{e}n), and fluting modes
(e.g.~\cite{ref28,ref29,ref30,ref31,ref32}); for the tube modes supported by
wide magnetic flux tubes, see
\cite{ref31}. Focusing on the longitudinal tube waves known to be an important
heating agent of solar magnetic regions, it is necessary to mention the recent
papers by
\cite{ref33}, which showed that the longitudinal flux tube waves are identified
as insufficient to heat the solar transition region and corona in agreement
with previous studies~\citep{ref34}.
\textbf{In other words, the problem of generation and transport of energy by
magnetic flux tubes remains unsolved in spite of its key role in physics of
various types of solar active regions.}

It is clear that this unsolved problem of energy transport by magnetic flux
tubes at the same time represents another unsolved problem related to the
energy transport and sunspot darkness (see 2.2 in \cite{Rempel2011}). From a
number of known concepts playing a noticeable role in understanding of the
connection between the energy transport and sunspot darkness, let us consider
the most significant theory, according to our vision. It is based on the
Parker-Biermann cooling effect \citep{ref41-3,Biermann1941,ref43-3} and
originates from the works of~\cite{Biermann1941} and~\cite{Alfven1942}.

\begin{figure}[tb]
\begin{center}
\includegraphics[width=12cm]{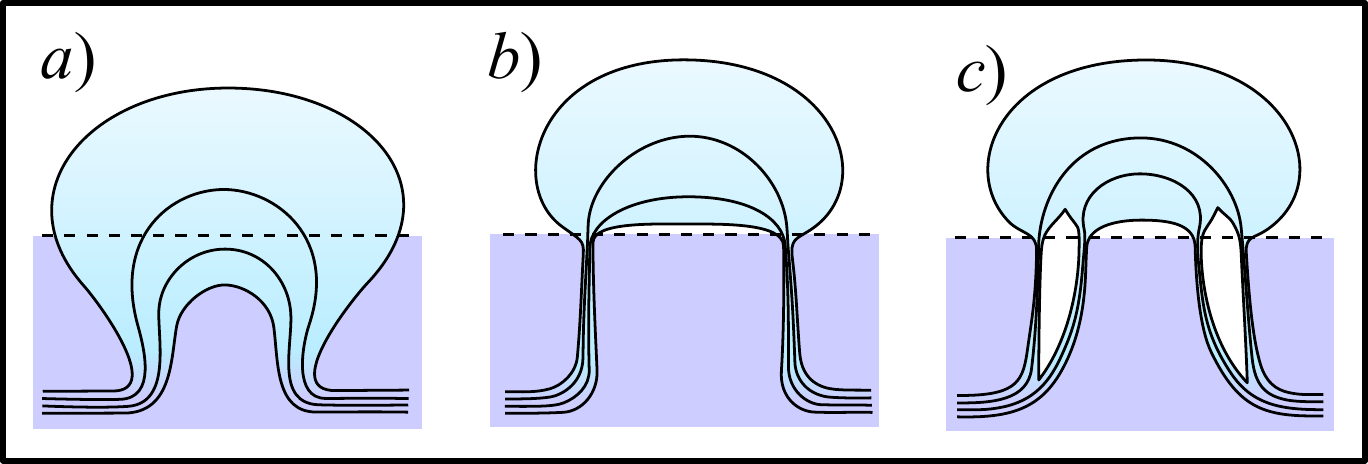}
\end{center}
\caption{The possible ways of a toroidal magnetic flux tube development into a
sunspot.
(a) A rough representation of the form a tube can take after the rise to the
surface by magnetic buoyancy (adopted from Fig.~2a in \cite{ref41-3});
(b) demonstrates the ``crowding'' below the photosphere surface because of
cooling (adopted from Fig.~2b in \cite{ref41-3});
(c) demonstrates the tube splitting as a consequence of the inner region
cooling under the conditions when the tube is in the thermal disequilibrium
with its surroundings and the convective heat transfer is suppressed
\mbox{\citep{Biermann1941}} above $\sim 0.71 R_{Sun}$. This effect as well as
the mechanism of the neutral atoms appearance inside the magnetic tubes
are discussed further in the text (see \mbox{Fig.~\ref{fig-lampochka}a}).
Adopted from Fig.~2c in \cite{ref41-3}.
} 
\label{fig04-3}
\end{figure}

The main point of the Parker-Biermann cooling effect is that the classical
mechanism of the magnetic tubes buoyancy (e.g. Fig.~\ref{fig04-3}a,
\cite{ref41-3}), emerging as a result of the shear flows instability
development in the tachocline, should be supplemented with the following
results of the~\cite{Biermann1941} postulate and the theory developed by
\cite{ref41-3,ref43-3}: the electric conductivity in the strongly ionized
plasma may be so high that the magnetic field becomes frozen into plasma and
causes the split magnetic tube (Fig.~\ref{fig04-3}b,c) to cool inside.

Biermann understood that the magnetic field within the sunspots might itself be
a reason of their darkness. Around the sunspots, the heat is transported up to
the surface of the Sun by means of convection (see 2.2.1 in~\cite{Rempel2011}),
while~\cite{Biermann1941} noted that such transport is strongly inhibited by
the nearly vertical magnetic field within the sunspot, thereby providing a
direct explanation for the reduced temperature at the visible surface. Thus,
the sunspot is dark because it is cooler than its surroundings, and it is
cooler because the convection is inhibited underneath.

Still, the missing cause of a very high conductivity in strongly ionized
plasma, which would produce a strong magnetic field ``frozen'' into this
plasma, has been the major flaw of the so called~\cite{Biermann1941} postulate.

Let us show a solution to the known problem of the Parker-Biermann cooling
effect, which is defined by the nature of the very large poloidal magnetic
fields in the tachocline (determined by the thermomagnetic Ettingshausen-Nernst
effect) and provides the physical basis for the photon channeling conditions 
inside the magnetic flux tubes.

\subsubsection{The thermomagnetic Ettingshausen-Nernst effect and poloidal magnetic field in the tachocline}

For the dynamo theories of planetary, stellar and spiral galactic magnetism the
Coriolis force is of crucial importance. However, the assumed large solar
dynamo leads to very large magnetic fields ($\sim 5 \cdot 10^7$ gauss
\citep{Fowler1955,Couvidat2003}, not observed on the surface of the Sun. This
requires an explanation of how these fields are screened from reaching the
surface.

As is known~\citep{Schwarzschild1958}, the temperature dependence of the
thermonuclear reaction rate in the region of 10$^7$K goes in proportion to
T$^{4.5}$. This means there is a sharp boundary between a much hotter region
where most of the thermonuclear reactions occur and a cooler region where they
are largely absent~\citep{Winterberg2015}. This boundary between radiative and
convective zones is the tachocline. It is the thermomagnetic
Ettingshausen-Nernst
effect~\citep{Ettingshausen1886,Sondheimer1948,Spitzer1956,Kim1969} which by
the large temperature gradient in the tachocline between the hotter and cooler
region leads to large currents shielding the large magnetic field of the dynamo
\citep{Winterberg2015}.

Subject to a quasi-steady state characterized by a balance of the magnetic
field of the dynamo, in the limit of weak collision (the collision frequency
much less than the cyclotron frequency of positive ions), a thermomagnetic
current can be generated in a magnetized
plasma~\citep{Spitzer1962,Spitzer2006}. For a fully ionized gases plasma the
thermomagnetic Ettingshausen-Nernst effect leads to a current density given by
(see Eqs.~(5-49) in~\citep{Spitzer1962,Spitzer2006}):

\begin{equation}
\vec{j} _{\perp} = \frac{3 k n_e c}{2 B^2} \vec{B} \times \nabla T
\label{eq06-01}
\end{equation}
\noindent where $n_e$ is the electron number density, $B$ is the magnetic
field, and $T$ is the absolute temperature (K). With $n_e = \left[ Z / (Z+1)
\right] n$, where $n = n_e + n_i$, and $n_i = n_e / Z$ is the ion number
density for a $Z$-times ionized plasma, the following is obtained:
\begin{equation}
\vec{j} _{\perp} = \frac{3 k n c}{2 B^2} \frac{Z}{Z+1} \vec{B} \times \nabla
T\, . \label{eq06-02}
\end{equation}
It exerts a force on plasma, with the force density $F$ given by
\begin{equation}
\vec{F} = \frac{1}{c} \vec{j} _{\perp} \times \vec{B} =
\frac{3 n k}{2 B^2} \frac{Z}{Z+1} \left( \vec{B} \times \nabla T \right)
\times \vec{B}
\label{eq06-03}
\end{equation}
or with $\nabla T$ perpendicular to $\vec{B}$
\begin{equation}
\vec{F} = \frac{3 n k}{2} \frac{Z}{Z+1} \nabla T
\label{eq06-04}
\end{equation}
leading to the magnetic equilibrium condition (see Eqs.~(4-1)
in~\citep{Spitzer1962})
\begin{equation}
\vec{F} = \frac{1}{c} \vec{j} _{\perp} \times \vec{B} = \nabla p
\label{eq06-05}
\end{equation}
with $p = (\rho / m) kT = nkT$. And by equating~(\ref{eq06-04})
and~(\ref{eq06-05}),
\begin{equation}
\frac{3 n k}{2} \frac{Z}{Z+1} \nabla T = nk \nabla T + kT \nabla n
\label{eq06-06}
\end{equation}
\noindent or
\begin{equation}
a \frac{\nabla T}{T} + \frac{\nabla n}{n} = 0,
~~~ where ~~ a = \frac{2 - Z}{2(Z+1)} ,
\label{eq06-06a}
\end{equation}
\noindent we obtain the condition:
\begin{equation}
T ^a n = const .
\label{eq06-07}
\end{equation}

For a singly-ionized plasma with $Z=1$, one has
\begin{equation}
T ^{1/4} n = const .
\label{eq06-08}
\end{equation}

For a doubly-ionized plasma ($Z=2$) one has $n=const$. Finally, in the limit $Z
\rightarrow A$, one has $T^{-1/2}n = const$. Therefore, $n$ does not strongly
depend on $T$, unlike in a plasma of constant pressure, in which $Tn=const$. It
shows that the thermomagnetic currents may change the pressure distribution in
magnetized plasma considerably.

Taking a Cartesian coordinate system with $z$ directed along $\nabla T$, the
magnetic field in the $x$-direction and the Ettingshausen-Nernst current in the
$y$-direction, and supposing a fully ionized hydrogen plasma with $Z=1$ in the
tachocline, one has
\begin{equation}
{j} _{\perp} = {j} _y = - \frac{3 n k c}{4 B} \frac{dT}{dz}. 
\label{eq06-09}
\end{equation}

From Maxwell's equation $4 \pi \vec{j}_{\perp}/ c = curl \vec{B}$, one has
\begin{equation}
{j} _y = \frac{c}{4 \pi} \frac{dB}{dz}, 
\label{eq06-10}
\end{equation}
and thus by equating~(\ref{eq06-09}) and~(\ref{eq06-10}) we obtain:
\begin{equation}
2B \frac{dB}{dz} = -6 \pi k n \frac{dT}{dz}.
\label{eq06-11}
\end{equation}

\begin{figure}[tb]
\begin{center}
\includegraphics[width=12cm]{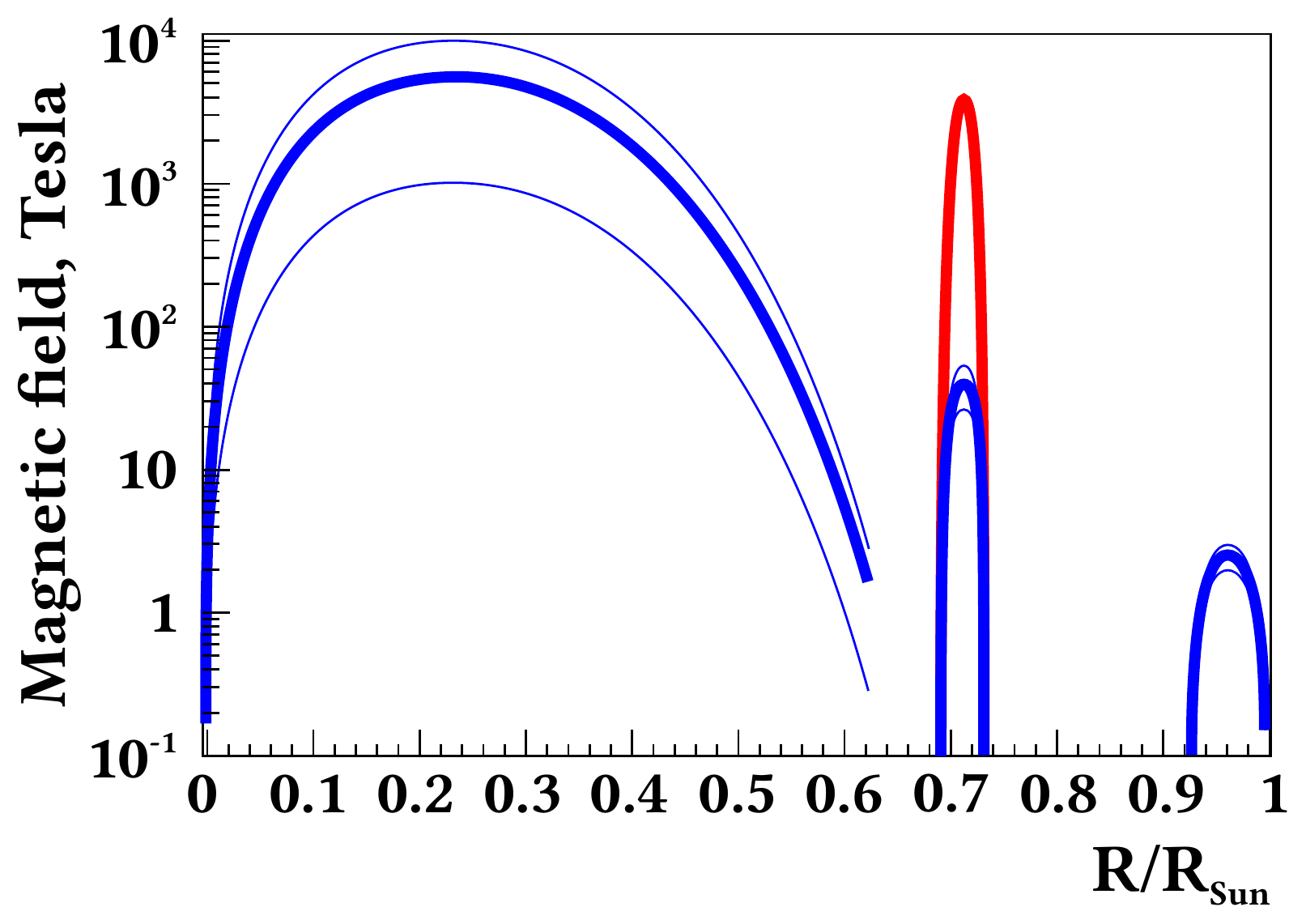}
\end{center}
\caption{The reconstructed solar magnetic field (in blue) simulation
from~\cite{Couvidat2003}: 10$^3$-10$^4$~Tesla (left), 30-50~Tesla (middle) and
2-3~Tesla (right), with a temperature of $\sim$9~MK, $\sim$2~MK
and~$\sim$200~kK, respectively. The thin lines show the estimated range of
values for each magnetic field component. Internal rotation was not included in
the calculation. An additional axion production at those places can modify both
intensity and shape of the solar axion spectrum (Courtesy Sylvaine 
Turck-Chi\`{e}ze (see Fig.~2 in~\cite{Zioutas2007})). The reconstructed solar 
magnetic field (in red) simulation from~(\ref{eq06-16}): $4 \cdot 10^3$~T in 
tachocline ($\sim0.7 R_{Sun}$).} 
\label{fig-R-MagField}
\end{figure}

From~(\ref{eq06-08}) one has
\begin{equation}
n = \frac{n _{OT} T_{OT}^{1/4}}{T^{1/4}},
\label{eq06-12}
\end{equation}
\noindent where the values $n = n_{OT}$ and $T = T_{OT}$ correspond to the
overshoot tachocline. Inserting~(\ref{eq06-12}) into~(\ref{eq06-11}), one finds
\begin{equation}
dB^2 = -\frac{6 \pi k n _{OT} T_{OT}^{1/4}}{T^{1/4}} dT,
\label{eq06-13}
\end{equation}
\noindent and hence, as a result of integration in the limits $[B_{OT},0]$ in
the left-hand side and $[0,T_{OT}]$ in the right-hand side,
\begin{equation}
\frac{B_{OT}^2}{8 \pi} = n _{OT} kT_{OT}
\label{eq06-14}
\end{equation}
which shows that the magnetic field of the thermomagnetic current in the
overshoot tachocline neutralizes the magnetic field of the dynamo reaching the
overshoot tachocline (see Fig.~\ref{fig-R-MagField}).

Hence, it is not hard to understand what forces compress the field into intense
filaments, in opposition to the enormous magnetic pressure
\begin{equation}
\frac{B_{OT}^2}{8 \pi} = p_{ext} \approx 6.5 \cdot 10^{13} \frac{erg}{cm^3} ~~
at ~~ 0.7 R_{Sun}, 
\label{eq06-15}
\end{equation}
\noindent where the gas pressure $p_{ext}$ at the tachocline of the Sun ($\rho
\approx 0.2 ~g\cdot cm^{-3}$ and $T \approx 2.3 \cdot 10^6 K$ \citep{ref45-3}
at~$0.7 R_{Sun}$) gives rise to a poloidal magnetic field

\begin{equation}
B_{OT} \simeq 4100 T. 
\label{eq06-16}
\end{equation}

According to (\ref{eq06-16}), a magnetic flux tube anchored in tachocline (see
Fig.~\ref{fig-twisted-tube}) has a significant toroidal magnetic field
($\sim$4100~T), within a layer near the base of the convection zone, where $0.7
R_{Sun}$ and $d \sim 0.05 R_{Sun}$ are constants defining the mean position and
thickness of the layer where the field is concentrated. Each of these anchored
magnetic flux tubes forms a pair of sunspots on the surface of the Sun.
Let us now show the theoretical possibility of the sunspot activity correlation
with the variations of the $\gamma$-quanta of axion origin, induced by the 
magnetic field variations in the overshoot tachocline.

\subsubsection{The Parker-Biermann cooling effect, Rosseland mean opacity and\\ axion-photon oscillations in twisted magnetic tubes}

\label{parker-biermann}

Several local models are known to have been used with great success to 
investigate the formation of buoyant magnetic transport, which transforms 
twisted magnetic tubes generated through shear amplification near the base 
tachocline (e.g.~\cite{Nelson2014}) and the structure and evolution of 
photospheric active regions (e.g.~\cite{Rempel2011a}).

\begin{figure}[tb]
\begin{center}
\includegraphics[width=15cm]{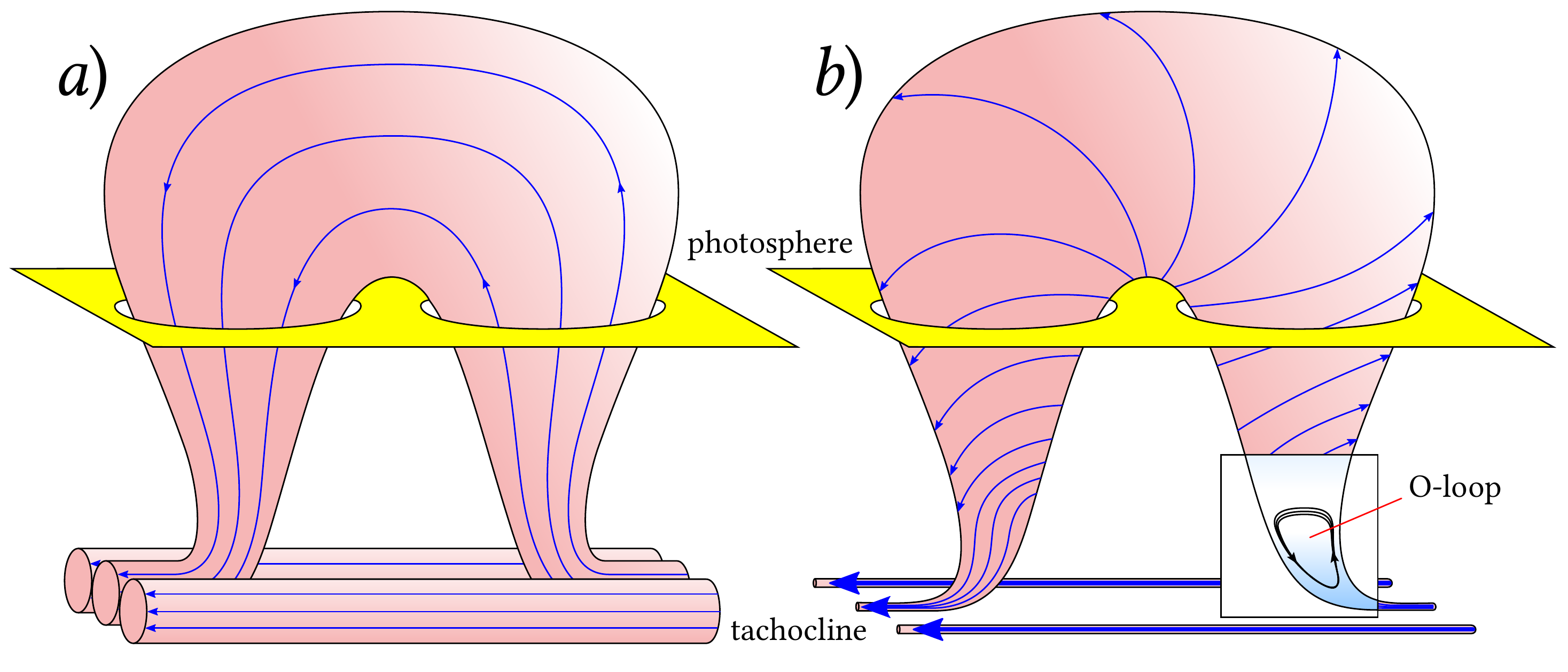}
\end{center}
\caption{An isolated and anchored in tachocline (a) magnetic flux tube (adopted
from~\cite{Parker1979}) and (b) twisted magnetic flux tube
(e.g.~\citep{Stein2012}, Fig.~2 in~\citep{Gold1960}, Fig.~1 and Fig.~2
in~\citep{Sturrock2001}) bursting through the solar photosphere to form a
bipolar region. \textbf{Inset in panel (b)}: topological effect of the magnetic
reconnection in the magnetic tube (see~\cite{Priest2000}), where the
$\Omega$-loop reconnects across its base, pinching off the $\Omega$-loop to
form a free $O$-loop (see Fig.~4 in~\cite{Parker1994}). The buoyancy of the
$O$-loop is limited by the magnetic tube interior with Parker-Biermann
cooling.} 
\label{fig-twisted-tube}
\end{figure}

Because these models assume the anchored magnetic flux tubes depending on the
poloidal field in the tachocline, it is not too hard to show that the magnetic
field $B_{OT}$ reaching $\sim 4100 T$ (see~(\ref{eq06-16})) may be at the same
time the reason for the Parker-Biermann cooling effect in the twisted magnetic
tubes (see~Fig.~\ref{fig-twisted-tube}b). The theoretical consequences of such
reasoning of the Parker-Biermann cooling effect are considered below.

First of all, we suggest that the classic mechanism of magnetic tubes buoyancy
(Fig.~\ref{fig-twisted-tube}а), appearing as a result of the shear instability
development in the tachocline, should be supplemented by the rise of the
twisted magnetic tubes in a stratified medium (Fig.~\ref{fig-twisted-tube}b
(see Fig.~1 and Fig.~2 in \citep{Sturrock2001}), where the magnetic field is
produced by dynamo action throughout the convection zone, primarily by
stretching and twisting in the turbulent downflows (see~\citep{Stein2012}).

Second, the twisting of the magnetic tube may not only promote its splitting,
but also may form a cool region under a certain condition 

\begin{equation}
p_{ext} = \frac{B^2}{8\pi}
\label{eq06v2-01}
\end{equation}

\noindent
when the tube (inset in~Fig.~\ref{fig-twisted-tube}b) is in the thermal
disequilibrium with its surroundings and the convective heat transfer is
suppressed \citep{Biermann1941}.

\begin{figure*}[tbp]
  \begin{center}
    \includegraphics[width=12cm]{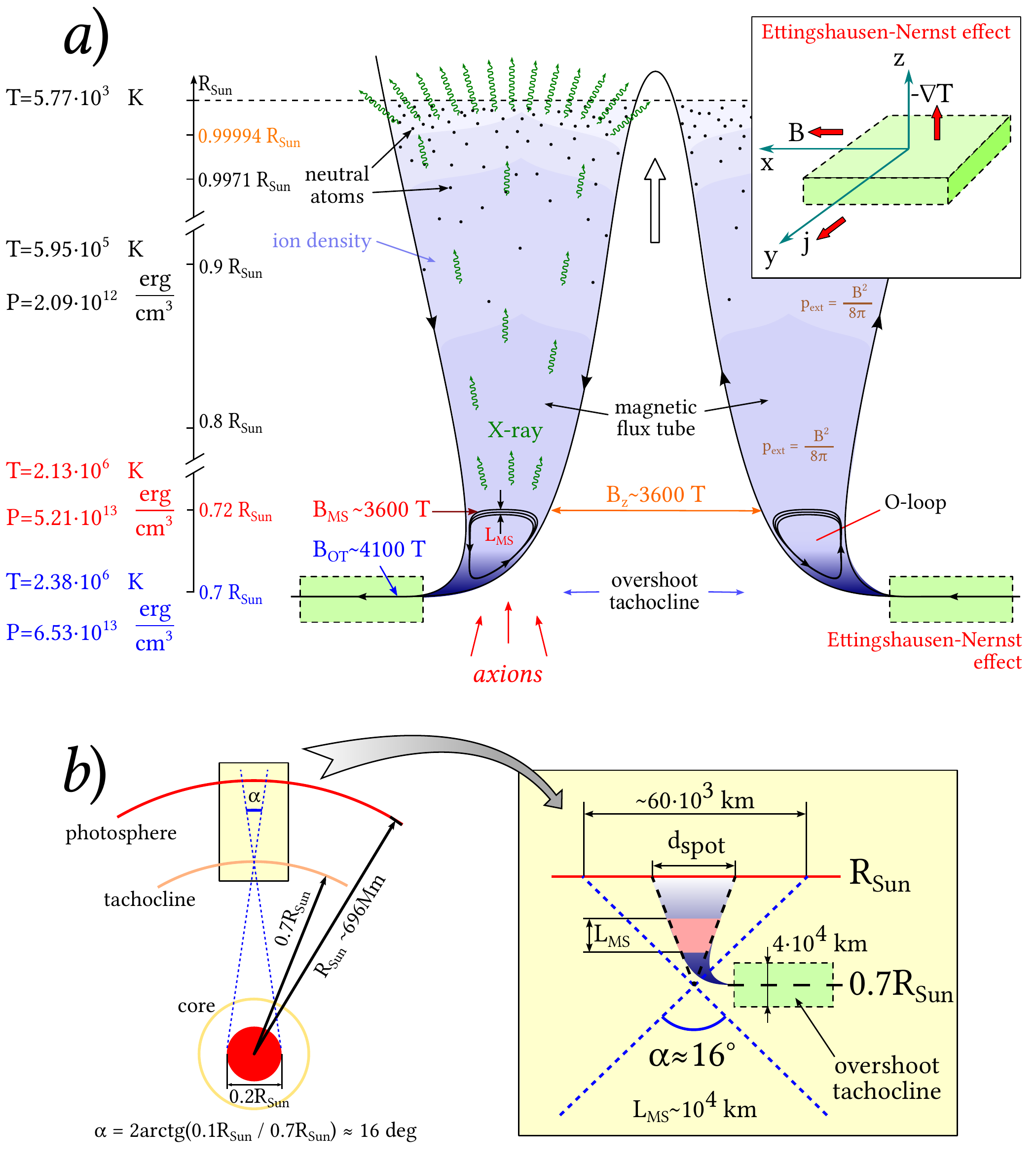}
  \end{center}
\caption{(a) Topological effects of the magnetic reconnection inside the
magnetic tubes with the ``magnetic steps''. The left panel shows the
temperature and pressure change along the radius of the Sun from the tachocline
to the photosphere \citep{ref45-3}, $L_{MS}$ is the height of the magnetic
shear steps. At $R \sim 0.72~R_{Sun}$ the vertical magnetic field reaches $B_z
\sim 3600$~T, and the magnetic pressure $p_{ext} = B^2 / 8\pi 
\simeq 5.21 \cdot 10^{13}~erg/cm^3$ \citep{ref45-3}. The very cool regions 
along the entire convective zone caused by the Parker-Biermann cooling effect 
have the magnetic pressure (\mbox{\ref{eq06v2-01}}) in the twisted magnetic tubes.
\newline (b) All the axion flux, born via the Primakoff effect (i.e. the real
thermal photons interaction with the Coulomb field of the solar plasma) comes
from the region  $\leq 0.1 R_{Sun}$~\citep{ref36}. Using the angle
$\alpha = 2 \arctan \left( 0.1 R_{Sun} / 0.7 R_{Sun} \right)$ marking the
angular size of this region relative to tachocline, it is possible to estimate
the flux of the axions distributed over the surface of the Sun. The flux of the
X-ray (of axion origin) is defined by the angle
$\gamma = 2 \arctan \left( 0.5 d_{spot} / 0.3 R_{Sun} \right)$, where
$d_{spot}$ is the diameter of a sunspot on the surface of the Sun (e.g.
$d_{spot} \sim 11000~km$~\citep{Dikpati2008}).}
\label{fig-lampochka}
\end{figure*}

It is interesting to explore how the cool region stretching from the 
tachocline to the photosphere, where the magnetic tube is in thermal 
non-equilibrium~(\ref{eq06v2-01}) with its surroundings, deals with the 
appearance of the neutral atoms (e.g. hydrogen) in the upper convection zone 
(see Fig.~\ref{fig-lampochka}a in contrast to Fig.~2c in \cite{Parker1955}). In
other words, how does this very cool region prevent the neutral atoms to 
penetrate from the upper convection zone to the base of the convection zone, 
i.e. tachocline?

It is essential to find the physical solution to the problem of solar convective zone which would fit the opacity experiments. The full calculation of solar opacities, which depend on the chemical composition, pressure and temperature of the gas, as well as the wavelength of the incident light, is a complex endeavour. The problem can be simplified by using a mean opacity averaged over all wavelengths, so that only the dependence on the gas physical properties remains (e.g. \cite{Rogers1994,Ferguson2005,Bailey2009}). The most commonly used is the Rosseland mean opacity $k_R$, defined as:

\begin{equation}
\frac{1}{k_R} = \left. \int \limits_{0}^{\infty} d \nu \frac{1}{k_\nu} \frac{dB_\nu}{dT} \middle/ 
\int \limits_{0}^{\infty} d \nu \frac{dB_\nu}{dT} \right.
\label{eq06v2-02}
\end{equation}

\noindent
where $dB_\nu / dT$ is the derivative of the Planck function with respect to 
temperature, $k_{\nu}$ is the monochromatic opacity at frequency $\nu$ of the 
incident light or the total extinction coefficient,  including stimulated 
emission plus scattering. A large value of the opacity indicates strong 
absorption from beam of photons, whereas a small value indicates that the beam 
loses very little energy as it passes through the medium.

\begin{figure}[tbp!]
\begin{center}
\includegraphics[width=15cm]{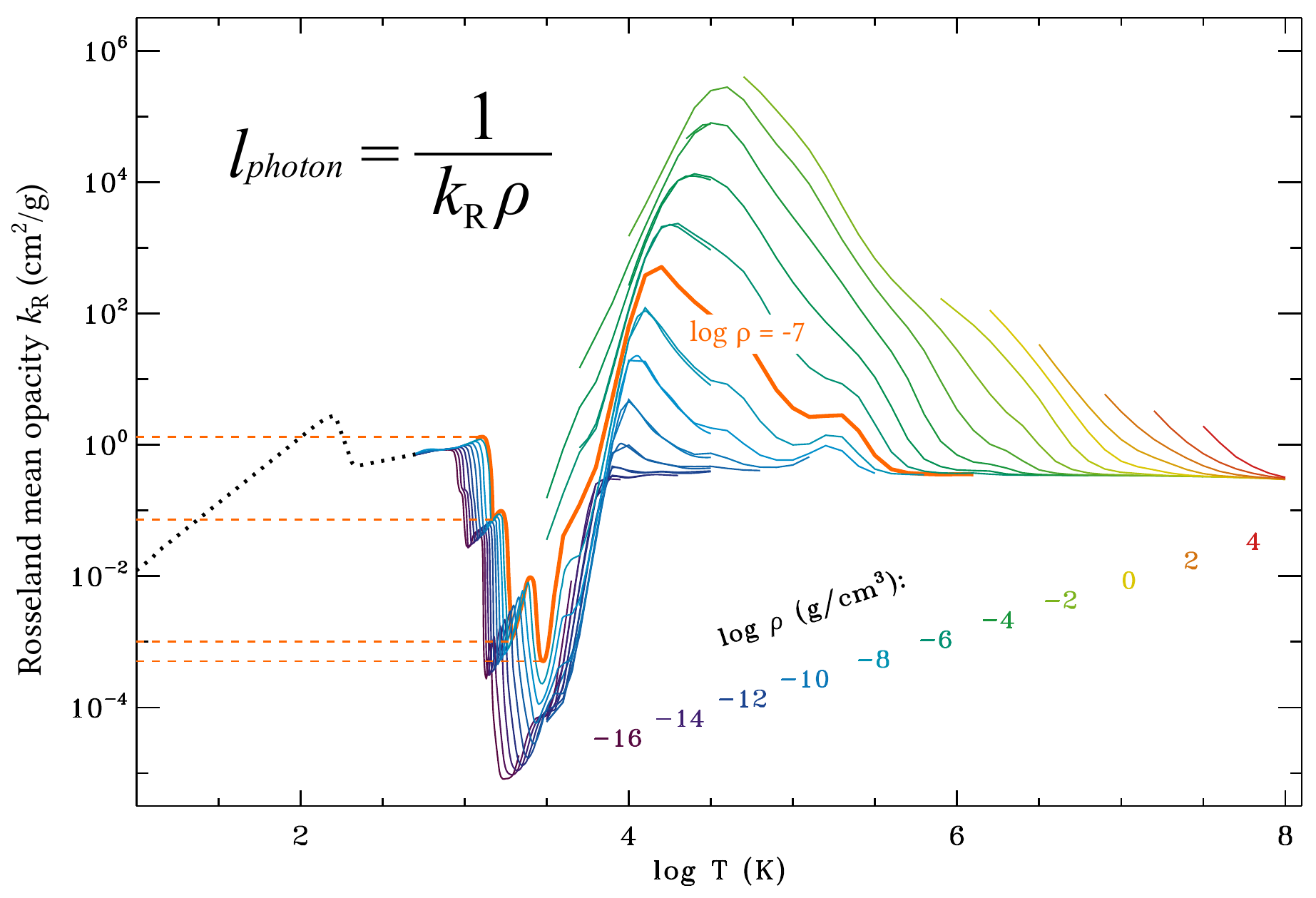}
\end{center}
\caption{Rosseland mean opacity $k_R$, in units of $cm^2 g^{-1}$, shown versus 
temperature (X-axis) and density (multi-color curves, plotted once per decade),
computed with the solar metallicity of hydrogen and helium mixture X=0.7 and 
Z=0.02. The panel shows curves of $k_R$ versus temperature for several 
``steady'' values of the density, labelled by the value of $\log {\rho}$ (in 
$g/cm^3$). Curves that extend from $\log {T} = 3.5$ to 8 are from the Opacity 
Project (opacities.osc.edu). Overlapping curves from $\log {T} = 2.7$ to 4.5 
are from \cite{Ferguson2005}. The lowest-temperature region (black dotted 
curve) shows an estimate of ice-grain and metal-grain opacity from 
\cite{Stamatellos2007}. Adapted from \cite{Cranmer2015}.}
\label{fig-opacity}
\end{figure}

Note that the Rosseland opacity is an harmonic mean, in which the greatest 
contribution comes from the lowest values of opacity, weighted by a function 
that depends on the rate at which the blackbody spectrum varies with 
temperature (see Eq.~(\ref{eq06v2-02}) and Fig.~\ref{fig-opacity}), and the
photons are most efficiently transported through the ``windows'' where $k_\nu$ 
is the lowest (see Fig.2 in \cite{Bailey2009}).

Taking the Rosseland mean opacities shown in Fig.~\ref{fig-opacity}, one may 
calculate, for example, four consecutive cool ranges within the convective 
zone (Fig.~\ref{fig-lampochka}a), where the internal gas pressure $p_{int}$ is 
defined by the following values:

\begin{equation}
p_{int} = n k_B T, ~where~ 
\begin{cases}
T \simeq 10^{3.48} ~K, \\
T \simeq 10^{3.29} ~K, \\
T \simeq 10^{3.20} ~K, \\
T \simeq 10^{3.11} ~K, \\
\end{cases}
\rho = 10^{-7} ~g/cm^3
\label{eq06v2-03}
\end{equation}

Since the inner gas pressure~(\ref{eq06v2-03}) grows towards the tachocline so 
that

\begin{align}
p_{int} &(T = 10^{3.48} ~K) \vert _{\leqslant 0.85 R_{Sun}}  > 
p_{int} (T = 10^{3.29} ~K) \vert _{\leqslant 0.9971 R_{Sun}} > \nonumber \\
& > p_{int} (T = 10^{3.20} ~K) \vert _{\leqslant 0.99994 R_{Sun}} > 
p_{int} (T = 10^{3.11} ~K) \vert _{\leqslant R_{Sun}} ,
\label{eq06v2-04}
\end{align}

\noindent
it becomes evident that the neutral atoms appearing in the upper convection 
zone ($\geqslant 0.85 R_{Sun}$) cannot descend deep to the base of the 
convection zone, i.e. tachocline (see Fig.~\ref{fig-lampochka}a).

Therefore it is very important to examine the connection between the Rosseland 
mean opacity and axion-photon oscillations in twisted magnetic tube.



Let us consider the qualitative nature of the $\Omega$-loop formation and
growth process, based on the semiphenomenological model of the magnetic
$\Omega$-loops in the convective zone.

\vspace{0.3cm}

\noindent $\bullet$ A high concentration azimuthal magnetic flux 
($B_{OT} \sim 4100$~T, see Fig.~\ref{fig-lampochka}) in the overshoot 
tachocline through the shear flows instability development.

An interpretation of such link is related to the fact that helioseismology
places the principal rotation $\partial \omega / \partial r$ of the Sun in the
overshoot layer immediately below the bottom of the convective zone
\citep{Parker1994}. It is also generally believed that the azimuthal magnetic
field of the Sun is produced by the shearing $r \partial \omega / \partial r$
of the poloidal field $B_{OT}$ from which it is generally concluded that the
principal azimuthal magnetic flux resides in the shear layer
\citep{Parker1955,Parker1993}.

\vspace{0.3cm}

\begin{figure}[tb]
  \begin{center}
    \includegraphics[width=15cm]{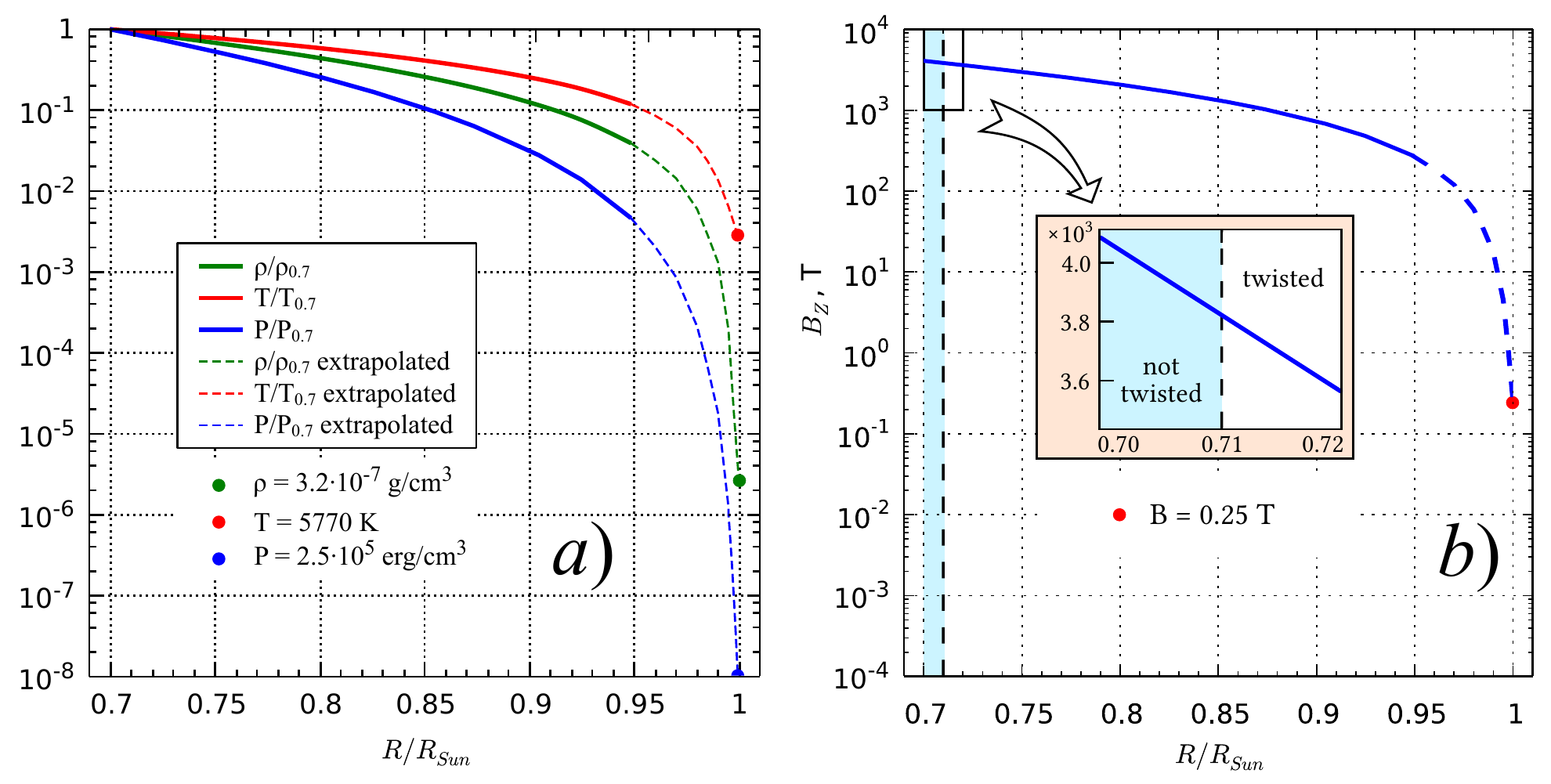}
  \end{center}
\caption{
(a) Normalized external temperature, density and gas pressure as functions of 
the solar depth $R/R_{Sun}$. The standard solar model with $He$ diffusion 
\citep{ref45-3} was used  for $R < 0.95 R_{Sun}$ (solid lines). The dotted 
lines mark extrapolated values.
(b) Variation of the magnetic field strength $B_z$  along the emerging
$\Omega$-loop as a function of the solar depth $R / R_{Sun}$ throughout the
convection zone. The solid blue line marks the permitted values for
the same standard solar model with $He$ diffusion \citep{ref45-3} starting at
the theoretical estimate of the magnetic field 
$B_{OT} \approx B_z(0) = 4100~T$. The dashed line is the continuation, 
according to the existence of the very cool regions inside the magnetic tube. 
Red point marks the up-to-date observations showing the mean magnetic field 
strength at the level $\sim 0.25~T = 2500 ~G$ \citep{Pevtsov2011,Pevtsov2014}.}
\label{fig-Bz}
\end{figure}

\noindent
$\bullet$ If some ``external'' factor of the local shear perturbation appears
against the background of the azimuthal magnetic flux concentration, such
additional local density of the magnetic flux may lead to the magnetic field
strength as high as, e.g. $B_z \sim 3600$~T (see Fig.~\ref{fig-lampochka}a
and \mbox{Fig.~\ref{fig-Bz}b}). Of course, this brings up a question about the 
physics behind such ``external'' factor and the local shear perturbation.

In this regard let us consider the superintense magnetic $\Omega$-loop
formation in the overshoot tachocline through the local shear caused by the
high local concentration of the azimuthal magnetic flux. The buoyant force
acting on the $\Omega$-loop decreases slowly with concentration so the vertical
magnetic field of the $\Omega$-loop reaches $B_z \sim 3600$~T at about 
$R / R_{Sun} \sim 0.72$ (see Fig.~\ref{fig-lampochka}a and Fig.~\ref{fig-Bz}b).
Because of the magnetic pressure (
see analog \mbox{(\ref{eq06-15})} and Fig.~\ref{fig-lampochka}a)
$p_{ext} = B_{0.72 R_{Sun}}^2 / 8\pi = 5.21\cdot 10^{13}~erg/cm^3$
\citep{ref45-3} this leads
to a significant cooling of the $\Omega$-loop tube (see
Fig.~\ref{fig-lampochka}a).

In other words, we assume the effect of the $\Omega$-loop cooling to be the
basic effect responsible for the magnetic flux concentration. It arises from
the well known suppression of convective heat transport by a strong magnetic
field~\citep{Biermann1941}. It means that although the principal azimuthal
magnetic flux resides in the shear layer, it predetermines the additional local
shear giving rise to a significant cooling inside the $\Omega$-loop.

Thus, the ultralow pressure is set inside the magnetic tube as a result of the
sharp limitation of the magnetic steps buoyancy inside the cool magnetic tube
(Fig.~\ref{fig-lampochka}a). This happens because the buoyancy of the magnetic
flows requires finite \textbf{superadiabaticity} of the convection zone
\citep{ref47-3,ref35-3}, otherwise, expanding according to the magnetic
\textbf{adiabatic} law (with the convection being suppressed by the magnetic
field), the magnetic clusters may become cooler than their surroundings, which
compensates the effect of the magnetic buoyancy of superintense magnetic O-loop.

Eventually we suppose that the axion mechanism based on the X-ray
channeling along the ``cool'' region of the split magnetic tube
\sethlcolor{pink}
\sethlcolor{yellow}
(Fig.~\ref{fig-lampochka}a) effectively supplies the necessary energy flux
``channeling'' in magnetic tube to the photosphere while the convective heat transfer is heavily
suppressed.

In this context it is necessary to have a clear view of the energy transport by the X-ray of axion origin, which are a primary transfer mechanism. The recent improvements in the calculation of the radiative properties of solar matter have helped to resolve several long-standing discrepancies between observations and the predictions of theoretical models (e.g. \cite{Rogers1994,Ferguson2005,Bailey2009}), and now it is possible to calculate the photon mean free path (Rosseland length) for Fig.~\ref{fig-opacity}:

\begin{equation}
l_{photon} = \frac{1}{k_R \rho} \sim 
\begin{cases}
2 \cdot 10^{10} ~cm  ~~ & for ~~ k_R \simeq 5 \cdot 10^{-4} ~cm^2/g, \\
10^{10} ~cm          ~~ & for ~~ k_R \simeq 10^{-3} ~cm^2/g, \\
1.5 \cdot 10^{8} ~cm ~~ & for ~~ k_R \simeq 6.7 \cdot 10^{-2} ~cm^2/g, \\
10^{7} ~cm           ~~ & for ~~ k_R \simeq 1 ~cm^2/g,
\end{cases}
~~ \rho = 10^{-7} ~g/cm^3
\label{eq06v2-05}
\end{equation}

\noindent
where the Rosseland mean opacity values $k_R$ and density $\rho$ are chosen so 
that the very low internal gas pressure $p_{int}$ (see Eq.~(\ref{eq06v2-04})) 
along the entire magnetic tube almost does not affect the external gas pressure
$p_{ext}$ (see (\ref{eq06v2-05}) and Fig.~\ref{fig-opacity}).

Let us now examine the appearance  of the X-ray of axion origin, induced by the magnetic field variations near the tachocline (Fig.~\ref{fig-lampochka}a) and their impact on the Rosseland length (see~(\ref{eq06v2-05})) inside the cool region of the magnetic tubes.

Let us remind that the magnetic field strength $B_{OT}$ in the overshoot 
tachocline of $\sim 4100~T$ (see Fig.~\ref{fig-lampochka}a) and the 
Parker-Biermann cooling effect in~(\ref{eq06v2-01}) lead to the corresponding 
value of the magnetic field strength $B(z = 0.72 R_{Sun}) \sim 3600 ~T$ 
(see Fig.~\ref{fig-lampochka}a), which in its turn assumes virtually zero 
internal gas pressure of the magnetic tube.

As it is shown above (see~\cite{Priest2000}), the topological effect of the
magnetic reconnection inside the $\Omega$-loop results in the formation of the
so-called O-loops (Fig.~\ref{fig-twisted-tube} and Fig.~\ref{fig-lampochka}a)
with their buoyancy limited from above by the strong cooling inside the
$\Omega$-loop (Fig.~\ref{fig-lampochka}a). It is possible to derive the value
of the horizontal magnetic field of the magnetic steps at the top of the O-loop:
$\vert B_{MS} \vert \approx \vert B(z = 0.72 R_{Sun}) \vert \sim 3600 ~T$.

So in the case of a large enough Rosseland lengh (see Eq.~(\ref{eq06v2-05})), 
X-ray of axion origin induced by the horizontal magnetic field in O-loops, reach
the photosphere freely, while in the photosphere itself, according to the 
Rosseland length

\begin{equation}
l_{photon} \approx 100 ~km < l \approx 300 \div 400 ~km,
\label{eq06v2-06}
\end{equation}

these photons undergo a multiple Compton scattering (see 
Section~\ref{subsec-osc-parameters}) producing a typical directional pattern
(Fig.~\ref{fig-lampochka}a).

Aside from the X-rays of axion origin with mean energy of 4.2~keV, there are 
only $h \nu \sim 0.95 ~keV$ X-rays (originating from the tachocline, according 
to a theoretical estimate by \cite{Bailey2009}) inside the magnetic tube. Such 
X-rays would produce the Compton-scattered photons with mean energy of 
$\leqslant 0.95~keV$ which contradicts the known measurements of the photons 
with mean energy of 3-4~keV (see Fig.~4 in \cite{Rieutord2014}). Our suggested 
theoretical model thus removes these contradictions by involving the X-rays of 
axion origin \textit{plus} the axions of the thermal X-ray origin, both 
produced in the magnetic field of O-loops (see Fig.~\ref{fig-lampochka}a and 
Fig.~\ref{app-b-fig01} in Appendix~\ref{appendix-luminosity}).

And finally, let us emphasize that we have just shown a theoretical possibility
of the time variation  of the  sunspot activity to correlate with the flux 
of the X-rays of axion origin; the latter being controlled by the magnetic 
field variations near the overshoot tachocline. As a result, it may be 
concluded that the the axion mechanism for solar luminosity variations 
based on the lossless X-ray ``channeling'' along the
magnetic tubes allows to explain the effect of the almost complete suppression of the
convective heat transfer, and thus to understand the known puzzling darkness of
the sunspots \citep{Rempel2011}.

\subsection{Estimation of the solar axion-photon oscillation parameters on the basis of the hadron axion-photon coupling in white dwarf cooling}

\label{subsec-osc-parameters}

It is known \citep{Cadamuro2012} that astrophysics provides a very interesting
clue concerning the evolution of white dwarf stars with their small mass
predetermined by the relatively simple cooling process. It is related to the
fact that recently it has been possible to determine their luminosity function
with the unprecedented precision \citep{Isern2008}. It seems that if the DFSZ
axion \citep{ref47,Dine1981} has a direct coupling to electrons and a decay
constant $f_a \sim 10^{9} ~GeV$, it provides an additional energy-loss channel
that permits to obtain a cooling rate that better fits the white dwarf
luminosity function than the standard one~\citep{Isern2008}. On the other hand,
the KSVZ axion \citep{ref46,ref46a}, i.e. the hadronic axion (with the mass in 
the $meV$ range and $g_{a\gamma \gamma} \sim 10^{-12} ~GeV^{-1}$) would also 
help in fitting the data, but in this case a stronger value for 
$g_{a\gamma \gamma}$ is required to perturbatively produce an electron coupling
of the required strength (\cite{Cadamuro2012}, Fig.~1 in \cite{Srednicki1985}, 
Fig.~1 in \cite{Turner1990}, Eq.~82 in \cite{Kim2010}).

Our aim is to estimate the solar axion-photon oscillation parameters basing on
the hadron axion-photon coupling derived from white dwarf cooling (see 
\mbox{Appendix~\ref{appendix-wd-cooling}}). The estimate
of the horizontal magnetic field in the O-loop is not related to the
photon-axion conversion in the Sun only, but also to the axions in the model of
white dwarf evolution. Therefore along with the values of the magnetic field
strength 
$B_{MS} \sim 3600 ~T$
and the height of the magnetic shear steps 
$L_{MS} \sim 1.28 \cdot 10^4 ~km$
(Fig.~\ref{fig-lampochka}a,b) we use the following parameters of the hadronic
axion (from the White Dwarf area in Fig.~\ref{fig05}a \citep{Irastorza2013,
Carosi2013}):

\begin{figure}[tbp!]
  \begin{center}
    \begin{minipage}[h]{0.44\linewidth}
      \includegraphics[width=7.6cm]{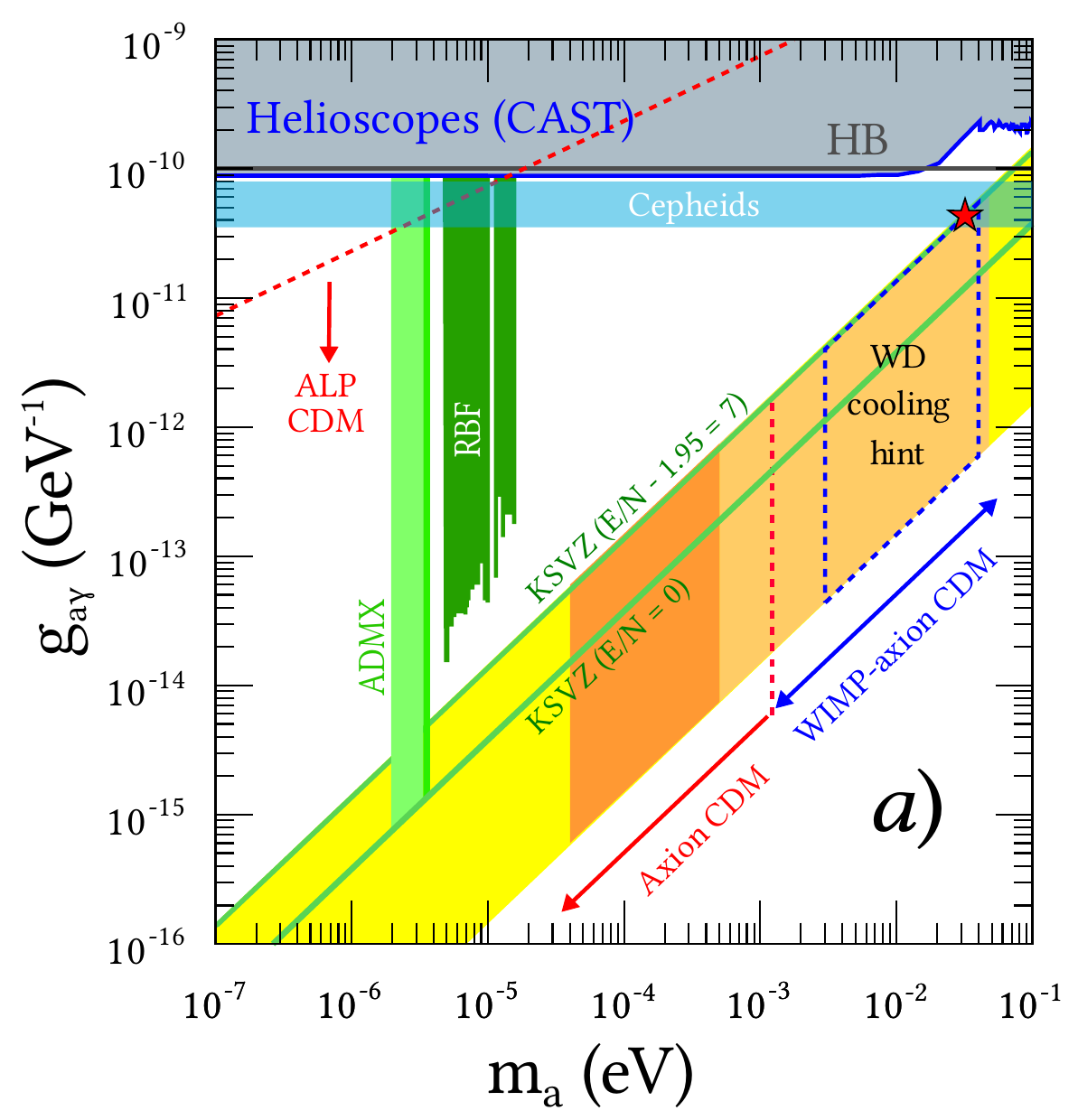}
    \end{minipage}
    \hfill
    \begin{minipage}[h]{0.53\linewidth}
      \includegraphics[width=8.8cm]{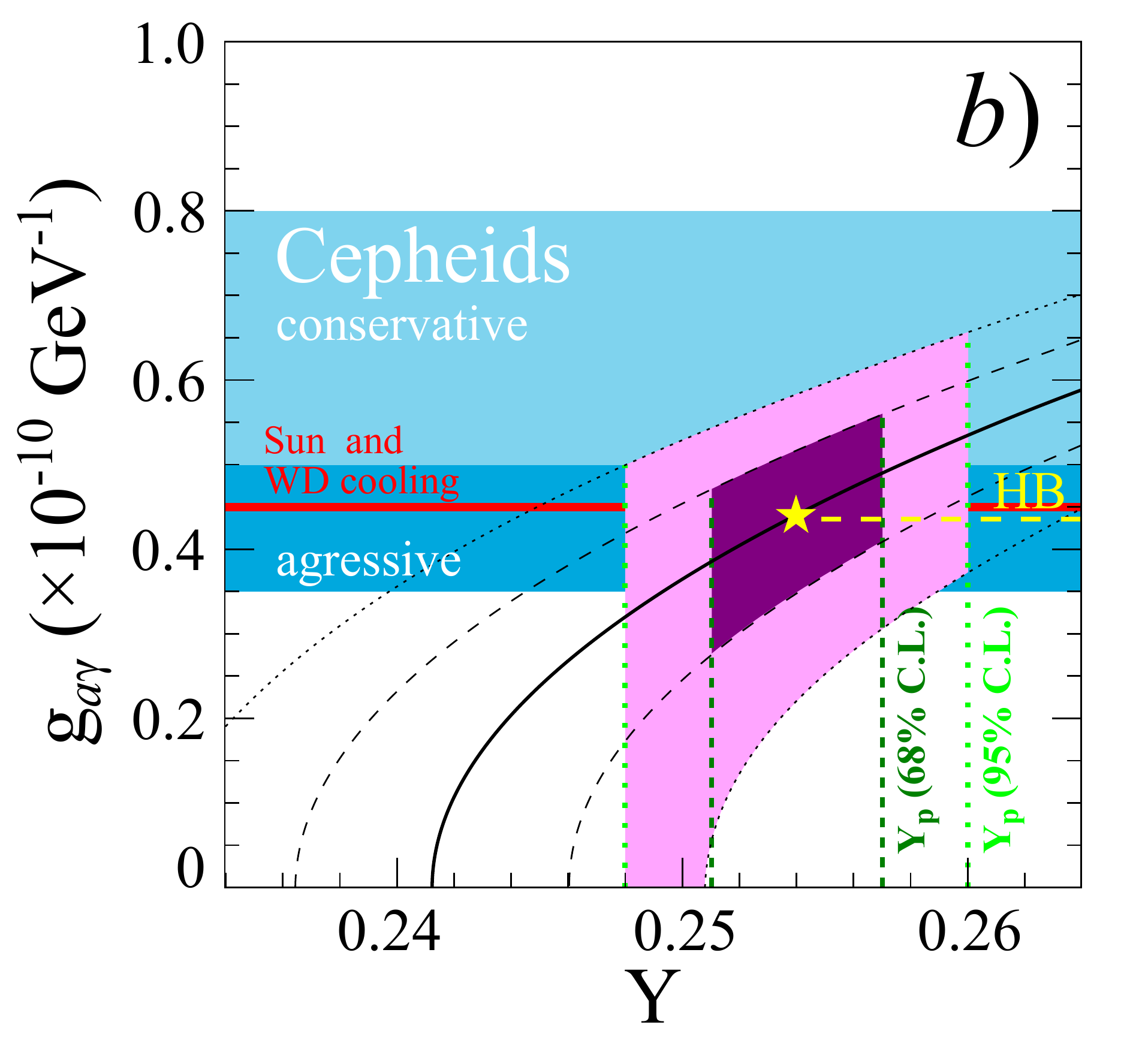}
    \end{minipage}
  \end{center}
\caption{\textbf{(a)} Summary of astrophysical, cosmological and laboratory
constraints on axions and axion-like particles. Comprehensive axion/ALP
parameter space, highlighting the two main front lines of direct detection
experiments: helioscopes (CAST~\citep{ref58,ref72,CAST2011,Arik2013}) and
haloscopes (ADMX~\citep{ref50} and RBF~\citep{ref51}). The astrophysical bounds
from horizontal branch and massive stars are labeled ``HB''~\citep{ref02} and
``Cepheids''~\citep{Carosi2013} respectively. The QCD motivated models
(KSVZ~\citep{ref46,ref46a} and DFSZ~\citep{ref47,Dine1981}) for axions lay in
the yellow diagonal band. The orange parts of the band correspond to
cosmologically interesting axion models: models in the ``classical axion
window'' possibly composing the totality of DM (labelled ``Axion CDM'') or a
fraction of it (``WIMP-axion CDM''~\citep{Baer2011}). For more generic ALPs,
practically all the allowed space up to the red dash line may contain valid ALP
CDM models~\citep{Arias2012}. The region of axion masses invoked in the WD
cooling anomaly is shown by the blue dash line~\citep{Irastorza2013}. The red
star marks the values of the axion mass $m_a \sim 3.2 \cdot 10^{-2} eV$ and the
axion-photon coupling constant $g_{a\gamma} \sim 4.4 \cdot 10^{-11} GeV^{-1}$
chosen in the present paper on the basis of the suggested relation between the
axion mechanisms of the Sun's and the white dwarf luminosity variations.
\newline
\textbf{(b)} $R$ parameter constraints to $Y$ and $g_{a \gamma}$ (adopted from
\cite{Ayala2014}. The dark purple area delimits the 68\%~C.L. for $Y$ and
$R_{th}$ (see Eq.~(1) in \cite{Ayala2014}). The resulting bound on the axion
($g_{10} = g_{a \gamma \gamma}/(10^{-10} ~GeV^{-1})$) is somewhere between a
rather conservative $0.5 < g_{10} \leqslant 0.8$ and most aggressive $0.35 <
g_{10} \leqslant 0.5$ \citep{Friedland2013}. The red line marks the values of
the axion-photon coupling constant $g_{a \gamma} \sim 4.4 \cdot 10^{-11}
~GeV^{-1}$ chosen in the present paper.
The blue shaded area represents the bounds from Cepheids
observation. The yellow star corresponds to $Y$=0.254 and the bounds from HB
lifetime (yellow dashed line).}
\label{fig05}
\end{figure}

\begin{equation}
g_{a \gamma} \sim 4.4 \cdot 10^{-11} ~ GeV^{-1}, ~~~ m_a \sim 3.2 \cdot 10^{-2} ~eV.
\label{eq3.30}
\end{equation}

The choice of these values is also related to the observed solar luminosity
variations in the X-ray band (see (\ref{eq3.35})). The theoretical
estimate and the
consequences of such choice are considered below.

%

As it is shown above, the $\sim 4100~T$ magnetic field in the overshoot
tachocline and the Parker-Biermann cooling effect in~(\ref{eq06v2-01}) may 
produce the O-loops with the horizontal magnetic field of
$\vert B_{MS} \vert \approx \vert B(z = 0.72 R_{Sun}) \vert \sim 3600 ~T$
stretching for about $L_{MS} \sim 1.28 \cdot 10^4 ~km$, and surrounded by virtually
zero internal gas pressure of the magnetic tube (see Fig.~\ref{fig-lampochka}a).

It is not hard to use the expression (\ref{eq08})
for the conversion probability\footnote{Hereinafter we use rationalized natural
units to convert the magnetic field units from $Tesla$ to $eV^2$, and the
conversion reads $1\,T = 195\,eV^2$~\citep{Guendelman2009}.}

\begin{equation}
P_{a \rightarrow \gamma} = \frac{1}{4} \left( g_{a \gamma} B_{MS} L_{MS} \right)^2 \sim 1
\label{eq3.31}
\end{equation}
for estimating the axion coupling constant to photons (\ref{eq3.30}).

Thus, it is shown that the hypothesis about the possibility for the solar
axions born in the core of the Sun to be efficiently converted back into
$\gamma$-quanta in the magnetic field of the magnetic steps of the O-loop
(above the solar overshoot tachocline) is relevant. Here the variations of the
magnetic field in the solar tachocline are the direct cause of the converted
$\gamma$-quanta intensity variations. The latter in their turn may be the cause
of the overall solar luminosity variations known as the active and quiet Sun phases.

It is easy to show that the theoretical estimate for the part of the axion
luminosity $L_a$ in the total luminosity of the Sun $L_{Sun}$ with respect to
(\ref{eq3.30}) is~\citep{ref58}

\begin{equation}
\frac{L_a}{L_{Sun}} = 1.85 \cdot 10 ^{-3} \left( 
\frac{g_{a \gamma}}{10^{-10} GeV^{-1}} \right)^2 \sim 3.6 \cdot 10^{-4} .
\label{eq3.32}
\end{equation}

\begin{figure*}
  \begin{center}
    \includegraphics[width=14cm]{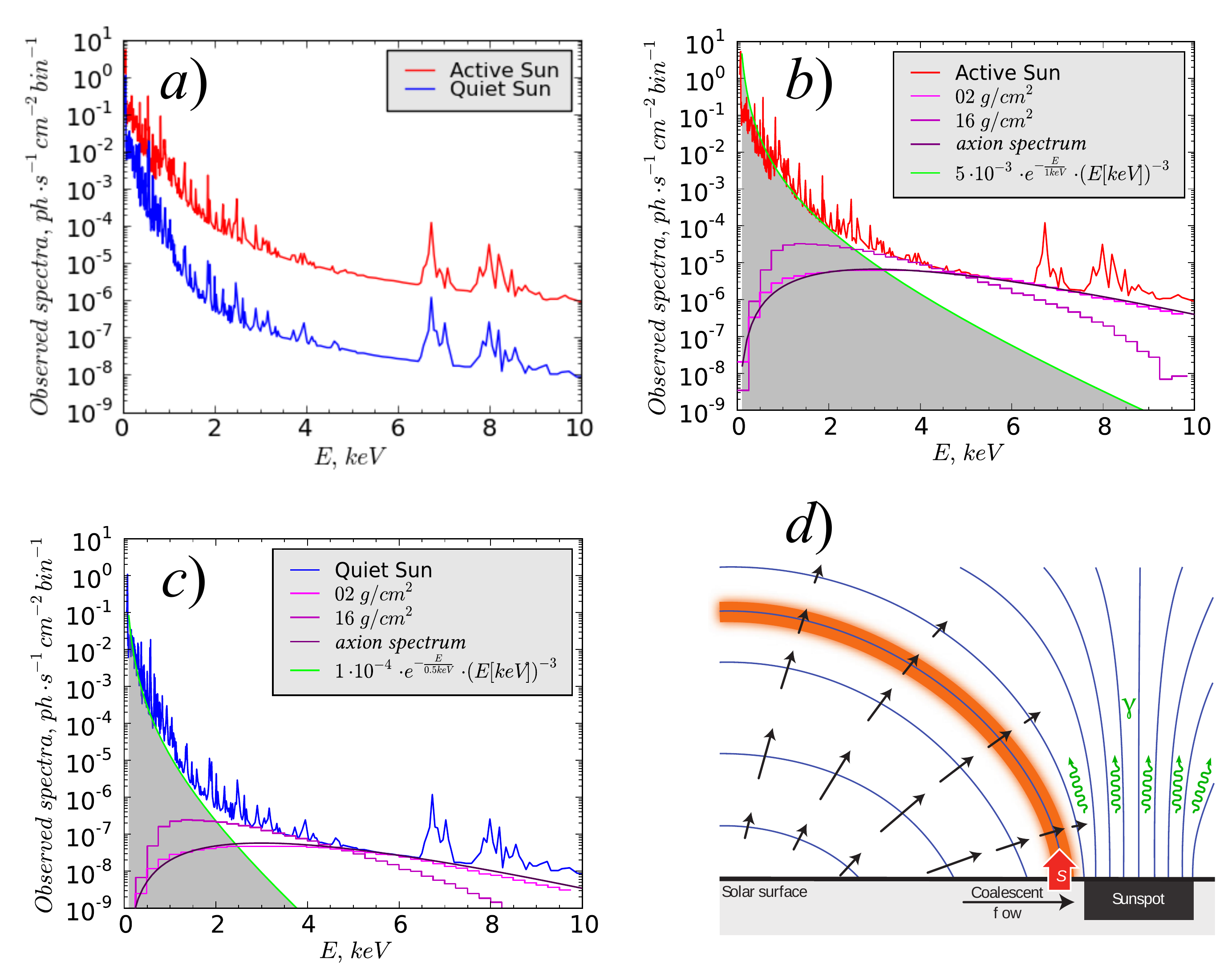}
  \end{center}

\caption{(a) Reconstructed solar photon spectrum below 10~keV from the active
Sun (red line) and quiet Sun (blue line) from accumulated observations
(spectral bin is 6.1~eV wide). Adopted from~\cite{ref59}.
\newline
(b) Reconstructed solar photon spectrum fit in the active phase of the Sun by
the quasi-invariant soft part of the solar photon spectrum (grey shaded area;
see \mbox{Eq.~(\ref{eq06-34})}) and three spectra (\ref{eq3.33}) degraded to
the Compton scattering for column densities above the initial conversion place
of 16 (adopted from~\cite{ref36}) and 2~$g / cm^2$ (present paper).
\newline
(c) The similar curves for the quiet phase of the Sun (grey shaded area 
corresponds to \mbox{Eq.~(\ref{eq06-35})})
\newline
(d) Cartoon showing the interplay between magnetic field expansion and the EUV
loop. A coalescent flow forming the sunspot drags the magnetic field in the
photosphere near the solar surface into the sunspot. In response, a hot spot of
enhanced upward directed Poynting flux, $S$, forms (red arrow). The expanding
field lines (blue) move upwards and to the side. When they transverse the hot
spot of Poynting flux, the plasma on that field line gets heated and brightens
up. As the field line expands further, it leaves the hot spot and gets darker
again. In consequence a bright coronal EUV loop forms (orange) and remains
rather stable as the successively heated field lines move through (adopted from
\cite{Chen2015}). X-ray emission is the $\gamma$-quanta of axion origin coming
from the magnetic tubes and not related to the magnetic reconnection as
conjectured by e.g. \cite{Shibata2011}.}
\label{fig06}
\end{figure*}

As opposed to the classic mechanism of the Sun modulation, the
axion mechanism is determined by the magnetic tubes rising to the photosphere,
and not by the over-photosphere magnetic fields. In this case the solar
luminosity modulation is determined by the axion-photon oscillations in the
magnetic steps of the O-loop causing the formation and channeling of the
$\gamma$-quanta inside the almost empty magnetic $\Omega$-tubes (see
Fig.~\ref{fig-twisted-tube} and Fig.~\ref{fig-lampochka}a). When the magnetic
tubes cross the photosphere, they ``open'' (Fig.~\ref{fig-lampochka}a), and the
$\gamma$-quanta are ejected to the photosphere, where their comfortable journey
along the magnetic tubes (without absorption and scattering) ends. As the
calculations by \cite{ref36} show, the further destiny of the $\gamma$-quanta
in the photosphere may be described by the Compton scattering, which actually
agrees with the observed solar spectral shape (Fig.~\ref{fig06}b,c).

From the axion mechanism point of view it means that the solar spectra during
the active and quiet phases (i.e. during the maximum and minimum solar
activity) differ from each other by the smaller or larger part of the Compton
spectrum, the latter being produced by the $\gamma$-quanta of the axion origin
ejected from the magnetic tubes into the photosphere (see Fig.~4 in
\cite{Chen2015}).

A natural question arises at this point: ``What are the real parts of the
Compton spectrum of the axion origin in the active and quiet phases of the Sun,
and do they agree with the experiment?'' Let us perform the
mentioned estimations basing on the known experimental results by ROSAT/PSPC,
where the Sun's coronal X-ray spectra and the total luminosity during the
minimum and maximum of the solar coronal activity were obtained~\citep{ref59}.

Apparently, the solar photon spectrum below 10~keV of the active and quiet Sun
(Fig.~\ref{fig06}a) reconstructed from the accumulated ROSAT/PSPC observations
may be described by three Compton spectra for different column densities rather
well (Fig.~\ref{fig06}b,c). This gives grounds for the assumption that the hard
part of the solar spectrum is mainly determined by the axion-photon conversion
efficiency:
\begin{align}
\left( \frac{d \Phi}{dE} \right)^{(*)} \simeq
\left( \frac{d \Phi}{dE} \right)^{(*)}_{corona} +
\left( \frac{d \Phi _{\gamma}}{dE} \right)^{(*)}_{axions} ,
\label{eq06-33}
\end{align}
\noindent where $\frac{d \Phi}{dE}$ is the observed solar spectra during the
active (red line in Fig.~\ref{fig06}a,b) and quiet (blue line in
Fig.~\ref{fig06}a,c) phases, $\left( \frac{d \Phi}{dE} \right)_{corona}$
represents the power-like theoretical solar spectra

\begin{equation}
\left( \frac{d \Phi}{dE} \right)_{corona} \sim E^{-(1+\alpha)} e^{-E/E_0} ,
\label{eq06-33a}
\end{equation}

\noindent
where a power law decay with the ``semi-heavy tail'' takes place in practice 
\citep{Lu1993} instead of the so-called power laws with heavy tails 
\citep{Lu1991,Lu1993} (see e.g. Figs.~3 and~6 in \cite{Uchaikin2013}). 
Consequently, the observed corona spectra 
($0.25 ~keV < E \leqslant 2.5 ~keV$) (shaded area in Fig.~\ref{fig06}b)

\begin{align}
\left( \frac{d \Phi}{dE} \right)^{(active)}_{corona} \sim
5 \cdot 10^{-3} \cdot (E~[keV])^{-3} \cdot \exp{\left(-\frac{E}{1 keV} \right)} 
~~for~the~active~Sun
\label{eq06-34}
\end{align}

\noindent and (shaded area in Fig.~\ref{fig06}c)

\begin{align}
\left( \frac{d \Phi}{dE} \right)^{(quiet)}_{corona} \sim
1 \cdot 10^{-4} \cdot (E~[keV])^{-3} \cdot \exp{\left(-\frac{E}{0.5 keV} \right)} 
~~for~the~quiet~Sun ;
\label{eq06-35}
\end{align}

\noindent $\left( \frac{d \Phi _{\gamma}}{dE} \right)_{axions}$ is the
reconstructed solar photon spectrum fit ($0 ~keV < E < 10 ~keV$) constructed
from three spectra (\ref{eq3.33}) degraded to the Compton scattering for
different column densities (see Fig.~\ref{fig06}b,c for the active and quiet
phases of the Sun respectively).

As is known, this class of flare models (Eqs.~(\ref{eq06-34})
and~(\ref{eq06-35})) is based on the recent paradigm in statistical physics
known as self-organized criticality
\citep{Bak1987,Bak1988,Bak1989,Bak1996,Aschwanden2011}. The basic idea is that
the flares are a result of an ``avalanche'' of small-scale magnetic
reconnection events cascading \citep{Lu1993,Charbonneau2001,Aschwanden2014} 
through the highly intense coronal magnetic structure \citep{Shibata2011} driven
at the critical state by the accidental photospheric movements of its magnetic
footprints. Such models thus provide a natural and computationally convenient
basis for the study of Parker hypothesis of the coronal heating by nanoflares
\citep{Parker1988}.

Another significant fact discriminating the theory from practice, or rather 
giving a true understanding of the measurements against some theory, should be 
recalled here (e.g. (\ref{eq06-33a}) (see Eq.~(5) in \cite{Lu1993})). The 
nature of power laws is related to the strong connection between the consequent
events (this applies also to the ``catastrophes'', which  in turn gives rise to
a spatial nonlocality related to the appropriate structure of the medium (see 
page 45 in \cite{Uchaikin2013})). As a result, the ``chain reaction'', i.e. the
avalanche-like growth of perturbation with more and more resource involved,
leads to the heavy-tailed distributions. On the other hand, obviously, none of 
the natural events may be characterized by the infinite values of mean and 
variance. Therefore, the power laws like (\ref{eq06-33a}) are approximate and
must not hold for the very large arguments. It means that the power law decay
of the probability density rather corresponds to the average asymptotics, and
the ``semi-heavy tails'' must be observed in practice instead.

In this regard we suppose that the application of the power-law distributions 
with semi-heavy tails leads to a soft attenuation of the observed corona 
spectra (which are not visible above $E > 2 \div 3 ~keV$), and thus to a close
coincidence between the observed solar spectra and $\gamma$-spectra of axion 
origin (Fig.~\ref{fig06}). I.e.

\begin{equation}
\left( \frac{d \Phi}{dE} \right)^{(*)} \simeq
\left( \frac{d \Phi _{\gamma}}{dE} \right)^{(*)}_{axions} 
~~~ \text{for energies} ~~ E > 2 \div 3 ~keV.
\label{eq06-35a}
\end{equation}

It means that the physics of the formation and ejection of the $\gamma$-quanta 
above $2 \div 3 ~keV$ through the sunspots into corona is not related to the 
magnetic reconnection theory by e.g. \cite{Shibata2011} (Fig.~\ref{fig06}d),
and may be of the axion origin.

With this in mind, let us suppose that the part of the differential solar axion
flux at the Earth~\citep{ref58}
\begin{align}
\frac{d \Phi _a}{dE} = 6.02 \cdot 10^{10} \left( \frac{g_{a\gamma}}{10^{10} GeV^{-1}} \right)^2 E^{2.481} \exp \left( - \frac{E}{1.205} \right) ~~cm^{-2}
s^{-1} keV^{-1} ,
\label{eq3.33}
\end{align}
\noindent which characterizes the differential $\gamma$-spectrum of the axion
origin $d \Phi _{\gamma} / dE$
(see $[ d \Phi _{\gamma} / dE ]_{axions}$ in (\ref{eq06-33}) and 
(\ref{eq06-35a}))

\begin{align}
\frac{d \Phi _{\gamma}}{dE}  \cong P_{\gamma} \frac{d \Phi _{a}}{dE}
~~ cm^{-2} s^{-1} keV^{-1} \approx
6.1 \cdot 10^{-3} P_{\gamma} \frac{d \Phi _{a}}{dE}
~ ph\cdot cm^{-2} s^{-1} bin^{-1}
\label{eq3.34}
\end{align}
\noindent
where the spectral bin width is 6.1~eV (see Fig.~\ref{fig06}a);
the probability $P_{\gamma}$ describing the relative portion of $\gamma$-quanta
(of axion origin) channeling along the magnetic tubes may be defined, according
to~\cite{ref59}, from the observed solar luminosity variations in the X-ray
band, recorded in ROSAT/PSPC experiments (Fig.~\ref{fig06}):
$\left(L_{corona}^X \right) _{min} \approx 2.7
\cdot 10^{26} ~erg/s$ at minimum and
$\left( L_{corona}^X \right) _{max} \approx 4.7 \cdot 10^{27} ~erg/s$
at maximum,

\begin{equation}
P_{\gamma} = P_{a \rightarrow \gamma} \cdot \dfrac{\Omega \cdot (0.5 d_{spot})^2}
{(\tan \left( \alpha / 2 \right) \cdot 0.3 R_{Sun})^2} \cdot \Lambda_a
\approx 3.4 \cdot 10^{-3},
\label{eq3.35}
\end{equation}

\noindent directly following from the geometry of the system
(Fig.~\ref{fig-lampochka}b), where the conversion probability 
$P_{a \rightarrow \gamma} \sim 1$ (\ref{eq3.31}); 

\begin{equation}
\Omega = (I_{\gamma ~CZ} / I_0) \cdot (I_{\gamma ~photo} / I_{\gamma ~CZ})
\cdot (I_{\gamma ~corona} / I_{\gamma ~photo}) \approx 0.23
\end{equation}

\noindent
is the total relative intensity of
$\gamma$-quanta, where $(I_{\gamma ~CZ} / I_0) \sim 1$ is the relative
intensity of $\gamma$-quanta ``channeling'' through the 
magnetic tubes in the convective zone,
$I_{\gamma ~photo} / I_{\gamma ~CZ} = \exp {[-(\mu l)_{photo}]} \sim 0.23$ (see
Eq.~\ref{eq06-43}) is the relative intensity of the Compton-scattered
$\gamma$-quanta in the solar photosphere, and $I_{\gamma ~corona} / I_{\gamma
~photo} = \exp {[-(\mu l)_{corona}]} \approx 1$ (see Eq.~\ref{eq06-44})
is the relative intensity of the Compton-scattered $\gamma$-quanta in the solar
corona;
$d_{spot}$ is the measured diameter of the sunspot
(umbra) \citep{Dikpati2008,Gough2010}. Its size determines the relative portion
of the axions hitting the sunspot area. Further,

\begin{equation}
\dfrac{(0.5 d_{spot})^2}{(\tan \left( \alpha / 2 \right) \cdot 0.3 R_{Sun})^2} \cong 0.034,
\end{equation}


\noindent where 

\begin{equation}
0.5 d_{spot} = \left[ \frac{1}{\pi} \left( 
\frac{\langle sunspot ~area \rangle _{max}}{\left\langle N_{spot} \right\rangle _{max}} 
\right) \right] ^{(1/2)} \cong 5500~km,
\end{equation}

\noindent and the value $\Lambda_a$ characterizes the portion of the 
axion flux going through the total $(2\left\langle N_{spot}
\right\rangle_{max})$ sunspots on the photosphere:

\begin{equation}
\Lambda_a = \dfrac{\left( sunspot\ axion\ flux \right)}{(1/3)\left( total\ axion\ flux \right)} \approx 
\dfrac{2 \left\langle N_{spot} \right\rangle _{max} (\tan \left( \alpha / 2 \right) \cdot 0.3 R_{Sun})^2}{(4/3) R_{Sun} ^2} \sim 0.42 , 
\label{eq3.36}
\end{equation}

\noindent and $\left\langle N_{spot} \right\rangle _{max} \approx 150$ is the
average number of the maximal sunspot number, and 
$\langle sunspot ~area \rangle _{max} \approx 7.5 \cdot 10^9 ~km^2 \approx 2470 ~ppm ~of ~visible ~hemisphere$
is the sunspot area (over the visible
hemisphere~\citep{Dikpati2008,Gough2010}) for the cycle 22 experimentally
observed by the Japanese X-ray telescope Yohkoh (1991)~\citep{ref36}.

On the other hand, from the known observations (see~\cite{ref59} and
Appendix~\ref{appendix-luminosity})
\begin{equation}
\frac{(L_{corona}^X)_{max}}{L_{Sun}} \cong 1.22 \cdot 10^{-6},
\label{eq3.37}
\end{equation}
\noindent where $L_{Sun} = 3.8418 \cdot 10^{33} erg / s$ is the solar
luminosity~\citep{ref63}. Using the theoretical axion impact estimate
(\ref{eq3.32}), one can clearly see that the obtained value (\ref{eq3.35}) is
in good agreement with the observations (\ref{eq3.37}):

\begin{equation}
P_{\gamma} =  \left. \frac{(L_{corona}^X)_{max}}{L_{Sun}} \middle/ 
\frac{L_a}{L_{Sun}} \sim 3.4 \cdot 10^{-3} \right. ,
\label{eq3.38}
\end{equation}

\noindent derived independently.

In other words, if the hadronic axions found in the Sun are the same particles
found in the white dwarfs with the known strength of the axion coupling to
photons (see (\ref{eq3.30}) and Fig.~\ref{fig05}a,b),
it is quite natural that the independent observations give the same estimate of
the probability $P_{\gamma}$ (see (\ref{eq3.35}) and (\ref{eq3.38})). So the
consequences of the choice (\ref{eq3.30}) are determined by the independent
measurements of the average sunspot radius, the sunspot
number~\citep{Dikpati2008,Gough2010}, the model estimates of the horizontal
magnetic field and the height $L_{MS}$ of the magnetic steps (see
Fig.~\ref{fig-lampochka}), and the hard part of the solar photon spectrum
mainly determined by the axion-photon conversion efficiency, and the
theoretical estimate for the part of the axion luminosity $L_a$ in the total
luminosity of the Sun $L_{Sun}$ (\ref{eq3.38}).

\section{Axion mechanism of the solar Equator -- Poles effect}

\begin{figure*}
\centerline{\includegraphics[width=12cm]{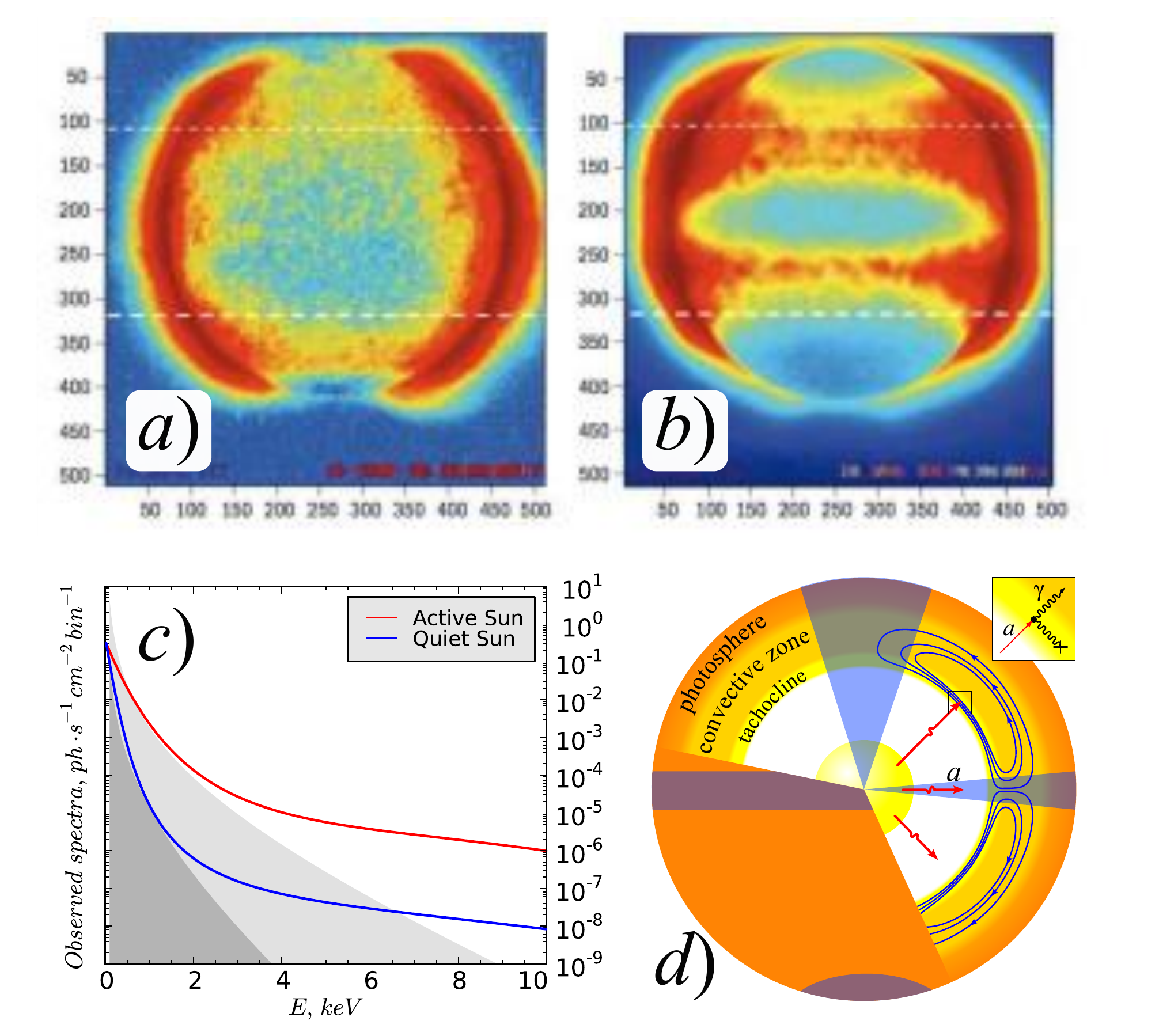}}

\caption{\textbf{Top:} Solar images at photon energies from 250~eV up to a few
keV from the Japanese X-ray telescope Yohkoh (1991-2001) (adopted
from~\cite{ref36}). The following is shown:
\newline
(a) a composite of 49 of the quietest solar periods during the solar minimum in 1996;
\newline
(b) solar X-ray activity during the last maximum of the 11-year solar cycle.
Most of the X-ray solar activity (right) occurs at a wide bandwidth of
$\pm 45^{\circ}$ in latitude, being homogeneous in longitude. Note that
$\sim$95\% of the solar magnetic activity covers this bandwidth.
\newline
\textbf{Bottom:} (c) Axion mechanism of solar irradiance variations 
above $2 \div 3 ~keV$, which is independent of the cascade reconnection
processes in corona (see shaded areas and Fig.~\ref{fig06}b,c,d),
and the red and blue curves characterizing the irradiance increment in the
active and quiet phases of the Sun, respectively;
\newline
(d) schematic picture of the radial travelling  of the axions inside the Sun.
Blue lines on the Sun designate the magnetic field. Near the tachocline 
(Fig.~\ref{fig-lampochka}a) the axions are converted into $\gamma$-quanta, 
which form the experimentally observed Solar photon spectrum after passing the 
photosphere (Fig.~\ref{fig06}). Solar axions that move towards the poles (blue 
cones) and in the equatorial plane (blue bandwidth) are not converted by 
Primakoff effect (inset: diagram of the inverse coherent process). The 
variations of the solar axions may be observed at the Earth by special 
detectors like the new generation CAST-helioscopes~\citep{ref68}. }
\label{fig-Yohkoh}
\end{figure*}

The axion mechanism of Sun luminosity variations is largely validated by the experimental
X-ray images of the Sun in the quiet (Fig.~\ref{fig-Yohkoh}a) and active
(Fig.~\ref{fig-Yohkoh}b) phases~\citep{ref36} which clearly reveal the
so-called Solar Equator -- Poles effect (Fig.~\ref{fig-Yohkoh}b).

The essence of this effect lies in the following. It is known that the axions
may be transformed into $\gamma$-quanta by inverse Primakoff effect in the
transverse magnetic field only. Therefore the axions that pass towards the
poles (blue cones in Fig.~\ref{fig-Yohkoh}b) and equator (the blue band in
Fig.~\ref{fig-Yohkoh}b) are not transformed into $\gamma$-quanta by inverse
Primakoff effect, since the magnetic field vector is almost collinear to the
axions' momentum vector. The observed nontrivial X-ray distribution in the
active phase of the Sun may be easily and naturally described within the
framework of the axion mechanism of the solar luminosity variations.

\begin{figure*}
\centerline{\includegraphics[width=15cm]{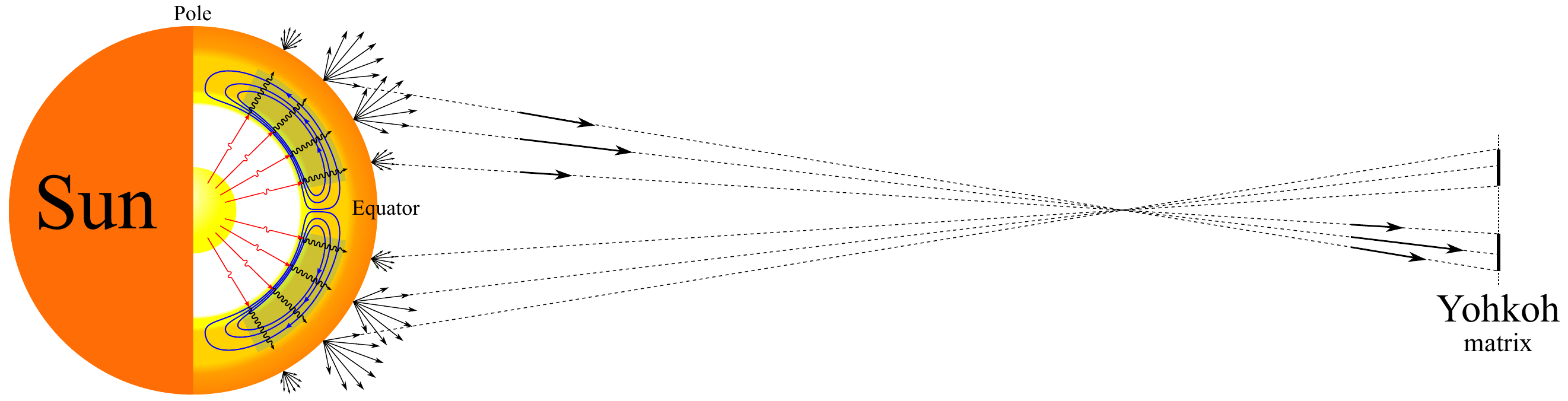}}

\caption{The formation of the high X-ray intensity bands on the Yohkoh
matrix. \label{fig07a}}
\end{figure*}

As described in Section~\ref{subsec-channeling}, the photons of axion origin 
travel through the convective zone along the magnetic flux tubes, up to the 
photosphere. In the photosphere they are Compton-scattered, which results in a 
substantial deviation from the initial axions directions of propagation 
(Fig.~\ref{fig07a}).

Let us make a simple estimate of the Compton scattering efficiency in terms of
the X-ray photon mean free path (MFP) in the photosphere:
\begin{equation}
l_{\mu} = (\mu)^{-1} = \left( \sigma_c \cdot n_e \right)^{-1} ,
\label{eq-compt-01}
\end{equation}
\noindent where 
$\mu$ is the total linear attenuation coefficient
(cm$^{-1}$),
the total Compton cross-section $\sigma_c = \sigma_0 = 8 \pi r_0^2 / 3$ for the
low-energy photons \citep{ref81,ref82}, $n_e$ is the electrons density in the
photosphere, and $r_0 = 2.8\cdot10^{-13}~cm$ is the so-called classical
electron radius.

Taking into account the widely used value of the matter density in the solar
photosphere $\rho \sim 10^{-7} ~g/cm^3$ and supposing that it consists of the
hydrogen (for the sake of the estimation only), we obtain that
\begin{equation}
n_e \approx \frac{\rho}{m_H}  \approx 6 \cdot 10^{16} ~ electron / cm^3\, ,
\label{eq-compt-02}
\end{equation}
which yields the MFP of the photon \citep{ref81,ref82}
\begin{align}
l_{\mu} = \left( 7 \cdot 10^{-25} ~cm^2 \cdot 6 \cdot 10^{16} ~electron/cm^3 \right)^{-1}
\approx  2.4 \cdot 10^7 ~cm = 240 ~km .
\label{eq-compt-03}
\end{align}

Since this value is smaller than the thickness of the solar photosphere
($l_{photo} \sim 300 \div 400 ~km$),
the Compton scattering is efficient
enough to be detected at the Earth 
(see \mbox{Fig.~\ref{fig07a}} and \mbox{Fig.~\ref{fig06}} adopted from
\mbox{\cite{ref59}});
\begin{equation}
\frac{I_{\gamma ~photo}}{I_{\gamma ~CZ}}
= \exp {\left[ - (\mu l)_{photo} \right]} \sim 0.23,
\label{eq06-43}
\end{equation}
\noindent which follows the particular case of the Compton scattering: Thomson
differential and total cross section for unpolarized photons
\citep{Griffiths1995}.

And finally taking into account that $l_{chromo} \sim 2 \cdot 10^3 ~km$ and
$n_e \sim 10^{13} ~electron/cm^3$ (i.e. $l_{\mu} \sim 1.4 \cdot 10^6 ~km$) and
$l_{corona} \sim 10^5 ~km$ and $n_e < 10^{11} ~electron/cm^3$ (i.e. $l_{\mu} >
1.4 \cdot 10^8 ~km$) (Fig.~12.9 in \cite{Aschwanden2004}), one may calculate
the relative intensity of the $\gamma$-quanta by Compton scattering in the
solar corona
\begin{equation}
\frac{I_{\gamma ~corona}}{I_{\gamma ~photo}}
= \frac{I_{\gamma ~chromo}}{I_{\gamma ~photo}} \cdot
\frac{I_{\gamma ~corona}}{I_{\gamma ~chromo}} =
\exp {\left[ - (\mu l)_{chromo} \right]} \cdot
\exp {\left[ - (\mu l)_{corona} \right]} \approx 1 ,
\label{eq06-44}
\end{equation}
\noindent which depends on the
total relative intensity of $\gamma$-quanta (see Eq.~(\ref{eq3.35})).

A brief summary is appropriate here. Coronal activity is a
collection of plasma  processes manifesting from the passage of magnetic fields
through it from below, generated by the solar dynamo in cycles of approximately
11 years (Fig.~\ref{fig-Yohkoh}). This global process culminating in the
reversal of the solar magnetic dipole at the end of each cycle involves the
turbulent dissipation of the magnetic energy, the flares  and heating of the
corona. The turbulent, highly dissipative, as well as largely ideal MHD
processes play their distinct roles, each liberating a comparable amount of
energy stored in the magnetic fields.

This mechanism is illustrated
in Fig.~\ref{fig06}d. When the magnetic flux erupts through the photosphere, it
forms a pair of sunspots pushing the magnetic field up and aside. The magnetic
field inside the sunspots is very high and the convection is suppressed.
Therefore the coalescence of the magnetic field is also suppressed. When some
magnetic field line crosses the region of the high Poynting flux, the energy is
distributed along this line in the form of plasma heating. This makes such line
visible in EUV band for some short time. While the magnetic field is being
pushed to the sides, next field line crosses the region of high Poynting flux
and flares up at the same position as the previous one and so on. This creates
an illusion of a static flaring loop, while the magnetic field is in fact
moving. It is interesting to note that \cite{Chen2015} expect the future
investigation to show to what extent this scenario also holds for X-ray
emission (see Supplementary Section~3 in \cite{Chen2015}).

In this context it is very important to consider the experimental observations
of solar X-ray jets (e.g. solar space missions of Yohkoh and Hinode
satellites), which show, for example, a gigantic coronal jet, ejecting from a
compact active region in a coronal hole \citep{Shibata1994} and tiny
chromospheric anemone jets \citep{Shibata2007}.

These jets are believed to be an indirect proof of small-scale ubiquitous
reconnection in the solar atmosphere and may play an important role in heating
it, as conjectured by Parker (\cite{Parker1988,Zhang2015,Sterling2015}).

Our main supposition here is that in contrast to EUV images (see orange line in
Fig.~\ref{fig06}d) and the coronal X-ray below $\sim 2 \div 3~keV$, the hard X-ray
emission above $\sim 3~keV$ is in fact the $\gamma$-quanta of axion origin,
born inside the magnetic tubes (see sunspot in Fig.~\ref{fig06}d), and is not
related to the mentioned indirect evidence (see e.g. Fig.~42 and Fig.~47 in
\cite{Shibata2011}) of the coronal jet, generated by the solar dynamo in cycles
of approximately 11 years (Fig.~\ref{fig-Yohkoh}). It will be interesting to
see if the proposed picture will ultimately be confirmed, modified, or rejected
by future observations and theoretical work to pin down the underlying physical
ideas.

Taking into account the directional patterns of the resulting radiation as well
as the fact that the maximum of the axion-originated X-ray radiation is
situated near 30 -- 40 degrees of latitude (because of the solar magnetic field
configuration), the mechanism of the high X-ray intensity bands formation on
the Yohkoh matrix becomes obvious. The effect of these bands widening near the
edges of the image is discussed in Appendix~\ref{appendix-widening} in detail.

\section{Summary and Conclusions}

In the given paper we present a self-consistent model of the axion mechanism of
the Sun's luminosity variations, in the framework of which we estimate the values of the axion
mass ($m_a \sim 3.2 \cdot 10^{-2} ~eV$) and the axion coupling constant to
photons ($g_{a \gamma} \sim 4.4 \cdot 10^{-11} ~GeV^{-1}$). A good
correspondence between the solar axion-photon oscillation parameters and the
hadron axion-photon coupling derived from white dwarf cooling (see
Fig.~\ref{fig05}) is demonstrated.

One of the key ideas behind the axion mechanism of Sun luminosity variations is the effect
of $\gamma$-quanta channeling along the magnetic flux tubes (waveguides inside
the cool region) above the base of the Sun convective zone
(Figs.~\ref{fig04-3}, \ref{fig-twisted-tube} and~\ref{fig-lampochka}). The low
refraction (i.e. the high transparency) of the thin magnetic flux tubes is
achieved due to the ultrahigh magnetic pressure (Fig.~\ref{fig-lampochka}a)
induced by the magnetic field of about 4100~T (see Eq.~(\ref{eq06-16}) and
Fig.~\ref{fig-lampochka}a). So it may be concluded that the axion mechanism of
Sun luminosity variations based on the lossless $\gamma$-quanta channeling along the
magnetic tubes allows to explain the effect of the partial suppression of the
convective heat transfer, and thus to understand the known puzzling darkness of
the sunspots (see 2.2.1 in \cite{Rempel2011}).

It is shown that the axion mechanism of luminosity variations (which means that
they are produced by adding the intensity variations of the $\gamma$-quanta of
the axion origin to the coronal part of the solar spectrum 
(Fig.~\ref{fig-Yohkoh}c)) easily explains the physics of the so-called Solar 
Equator -- Poles effect observed in the form of the anomalous X-ray 
distribution over the surface of the active Sun, recorded by the Japanese X-ray
telescope Yohkoh (Fig.~\ref{fig-Yohkoh}, top).

The essence of this effect consists in the following: axions that move towards
the poles (blue cones in Fig.~\ref{fig-Yohkoh}, bottom) and equator (blue
bandwidth in Fig.~\ref{fig-Yohkoh}, bottom) are not transformed into
$\gamma$-quanta by the inverse Primakoff effect, because the magnetic field
vector is almost collinear to the axions' momentum in these regions (see the
inset in Fig.~\ref{fig-Yohkoh}, bottom). Therefore the anomalous X-ray
distribution over the surface of the active Sun is a kind of a ``photo'' of the
regions where the axions' momentum is orthogonal to the magnetic field vector
in the solar over-shoot tachocline. The solar Equator -- Poles effect is not
observed during the quiet phase of the Sun because of the magnetic field
weakness in the overshoot tachocline, since the luminosity increment of the
axion origin is extremely small in the quiet phase as compared to the active
phase of the Sun.

In this sense, the experimental observation of the solar Equator -- Poles
effect is the most striking evidence of the axion mechanism of Sun luminosity
variations. It is hard to imagine another model or considerations which would
explain such anomalous X-ray radiation distribution over the active Sun surface
just as well (compare Fig.~\ref{fig-Yohkoh}a,b with Fig.~\ref{app-b-fig01}a
).

And, finally, let us emphasize one essential and the most painful point of the
present paper. It is related to the key problem of the axion mechanism of the solar
luminosity variations and is stated rather simply: ``Is the process of axion conversion
into $\gamma$-quanta by the Primakoff effect really possible in the magnetic
steps of an O-loop near the solar overshoot tachocline?'' This question is
directly connected to the problem of the hollow magnetic flux tubes existence
in the convective zone of the Sun, which are supposed to connect the tachocline
with the photosphere. So, either the more general theory of the Sun or the
experiment have to answer the question of whether there are the waveguides in
the form of the hollow magnetic flux tubes in the cool region of the convective
zone of the Sun, which are perfectly transparent for $\gamma$-quanta, or our
model of the axion mechanism of Sun luminosity variations is built around simply guessed
rules of calculation which do not reflect any real nature of things.

\section*{Acknowledgements}

\noindent The work of M. Eingorn was supported by NSF CREST award HRD-1345219
and NASA grant NNX09AV07A.

\appendix
\numberwithin{equation}{section}
\numberwithin{figure}{section}
\numberwithin{table}{section}

\

\section{Estimation of the hadron axion-photon coupling in white dwarf cooling}
\label{appendix-wd-cooling}

In order to estimate the parameters of the hadron axion-photon coupling

\begin{equation}
g_{a \gamma} \sim 4.4 \cdot 10^{-11} ~GeV^{-1}, ~~ m_a \sim 3.2 \cdot 10^{-2} ~eV
\label{eq-a2-01}
\end{equation}

\noindent
from white dwarf cooling, we focus on white dwarf stars (e.g. G117-B15A), which represent the final evolutionary stages of low- and intermediate-mass stars. Since they are strongly degenerate and do not have relevant nuclear energy sources, their evolution may be described as a slow cooling process in which the gravothermal energy release is the main energy source driving their evolution (e.g. \cite{Corsico2012}).

On the other hand, if there is dark matter in the Universe, some excellent candidates may exist in the galaxy and in the white dwarf stars in the form of weakly interacting particles, as it was recognized earlier e.g. by \cite{Raffelt1986} for the case of axions.

In particular, stars can be used to put constraints on the mass of the axion \citep{Raffelt1996}, where the coupling strength of KSVZ axions to electrons and to photons is defined through a dimensionless coupling constant $g_{ae}$, and dimension  coupling constant $g_{a \gamma}$.

For example, from the so called radiatively induced coupling to electrons (e.g. Fig. 1 in \cite{Srednicki1985}, Fig. 1 in \cite{Turner1990}
):

\begin{equation}
(g_{ae})_{KSVZ} = \frac{3 \alpha^2 m_e}{2 \pi f_a} \left[ \frac{E}{N} \ln \frac{f_a}{m_e} - \frac{2}{3} \frac{4 + z + w}{1 + z + w} \ln \frac{\Lambda_{QCD}}{m_e} \right] ,
\label{eq-a2-02}
\end{equation}

\noindent
where $\alpha \simeq 1/137$ is the fine structure constant, $m_e = 0.511 ~MeV$ is electron mass, $f_a$ is the energy scale of the spontaneous breaking of the Peccei-Quinn U(1) symmetry, $E/N$ is the ratio between the electromagnetic $E$ and color $N$ anomalies, the part proportional to $(4+z +w)/(1+z+w)$ arises from axion/pion mixing, and is cut off at the QCD confinement  scale $\Lambda_{QCD} \approx 200 ~MeV$, $z$ and $w$ are the quark mass ratios $m_u / m_d$ and $m_u / m_s$ introduced in equation of the axion mass

\begin{equation}
m_a = \frac{m_\pi f_\pi}{f_a} \sqrt{\frac{z}{(1+z+w)(1+z)}} \simeq 6 ~meV \frac{10^9 ~GeV}{f_a},
\label{eq-a2-03}
\end{equation}

\noindent
it is not difficult to derive

\begin{equation}
(g_{ae})_{KSVZ} \approx 8 \cdot 10^{-15},
\label{eq-a2-04}
\end{equation}

\noindent
where $m_\pi = 135 ~MeV$ is the pion mass and $f_{\pi} \simeq 92 ~MeV$ is the pion decay constant, $f_a \approx 1.85 \cdot 10^8 ~GeV$, typical theoretical models with $\left \vert E/N - 1.95 \right \vert = 7$ (see Fig.~\ref{fig05}a).

The energy loss rate due to radiatively induced coupling of the hadronic axion to electrons (instead of the axion bremsstrahlung \citep{Nakagawa1987,Nakagawa1988}) as (4) is written

\begin{equation}
\varepsilon _{e \rightarrow a} ^{WD} = 1.08 \cdot 10^{23} \frac{erg}{g \cdot s} \cdot \frac{(g_{ae}^2)_{KSVZ}}{4 \pi} \cdot \frac{Z^2}{A} \cdot T_7 ^4 \cdot F(T, \rho),
\label{eq-a2-05}
\end{equation}

\noindent
where $Z$ and $A$ are the atomic and mass numbers of the plasma components, 
$T_7 = T/10^7 ~K$, $F$ is a function of the temperature and the density which 
takes into account the properties of the plasma, and is of order unit throughout
most of the interior of a typical white dwarf model (e.g. 
\cite{BischoffKim2008}).

Consequently, it may be shown (see Eq.5 in \cite{BischoffKim2008}) that the 
fraction of axion luminosity is then

\begin{equation}
L_{e \rightarrow a}^{WD} = \varepsilon _{e \rightarrow a}^{WD} \cdot M_* \simeq 1.03 \cdot 10^{57} \cdot \frac{(g_{ae}^2)_{KSVZ}}{4\pi} ~erg \cdot s^{-1},
\label{eq-a2-06}
\end{equation}

\noindent
leading to the theoretical estimation of the axion fraction in the white dwarfs 
(stellar mass $M_* = 0.55 M_{Sun}$, effective temperature $T_{eff} = 12000~K$ 
(established spectroscopically in \cite{Corsico2012}) and Sun luminosity

\begin{equation}
\log (L_{e \rightarrow a}^{WD} / L_{Sun}) \approx -5.8 .
\label{eq-a2-07}
\end{equation}

\begin{figure}[tb]
  \begin{center}
    \includegraphics[width=15cm]{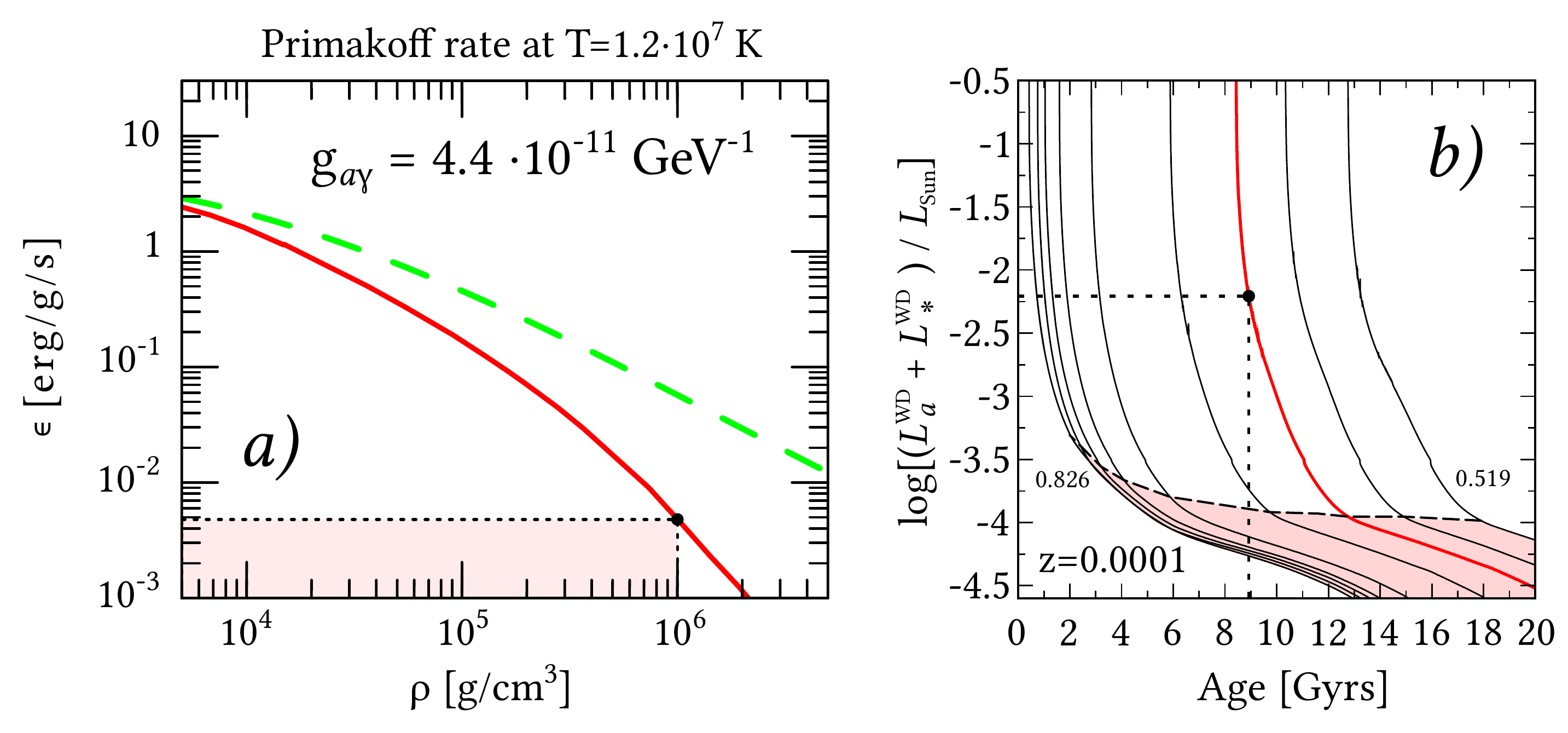}
  \end{center}
\caption{(a) Energy-loss rate of a helium plasma at $T = 1.2 \cdot 10^7 ~K$  of 
the pulsating white dwarf star G117−B15A by axion emission with $g_{a\gamma} = 
4.4 \cdot 10^{-11} ~GeV^{-1}$. The solid line is from transverse-longitudinal 
fluctuations (degenerate, nonrelativistic); the dashed line is the corresponding
classical (nondegenerate, nonrelativistic) limit (see Eq.(5.9) in \cite{Raffelt1996}). (Adapted from \cite{Altherr1994}).
(b) Total luminosity as a function of age for a set of sequences with four 
different metallicity of the progenitor star. The age corresponds to the total 
age of the model, by considering all the stages previous to the white dwarf 
cooling. The horizontal dashed line represents the point in the evolution where 
the crystallization process begins at the core, by means of the 
\cite{Horowitz2010} phase diagram for crystallization. The stellar mass values 
in solar mass units, from right to left, are: $M_* / M_{Sun} = 0.519, 0.534, \textcolor{red}{0.550}, 0.561,
0.569, 0.621, 0.669, 0.708, 0.737 ~and~ 0.826$ for $Z = 0.0001$. (Adapted from 
\cite{Romero2015}).}
\label{fig-a2-01}
\end{figure}

Meanwhile, using the coupling constant $g_{a\gamma}$ (\ref{eq-a2-01}), as in 
Fig.~\ref{fig-a2-01}a, it is evident that the hadron axion emission rate is a 
steeply falling function of density $\rho = 10^6 ~g/cm^3$ when degeneracy 
effects become important (see Eq.(5.9) in \cite{Raffelt1996}).

Thus the total axion luminosity is

\begin{equation}
\log \frac{L_a ^{WD}}{L_{Sun}} = \log \frac{L_{e \rightarrow a}^{WD} + L_{\gamma \rightarrow a} ^{WD}}{L_{Sun}} \simeq \log \frac{L_{\gamma \rightarrow a}^{WD}}{L_{Sun}} = -2.3
\label{eq-a2-08}
\end{equation}

\noindent
where

\begin{equation}
\frac{L_{\gamma \rightarrow a}^{WD}}{L_{Sun}} \sim 4.7 \cdot 10^{-3},
\label{eq-a2-08a}
\end{equation}

\noindent
for pulsating white dwarfs (e.g. G117-B15A with  temperature $T \simeq 1.2 \cdot
10^7 ~K$ and density $\rho = 10^6 ~g/cm^3$ of isothermal core 
\citep{Corsico2001} and the radius $R = 9.6 \cdot 10^8 ~cm$ \citep{Kepler2000}, 
which  yields (\ref{eq-a2-03})) is found using the Raffelt-DKTW-APG equation 
(see Fig.2 in \cite{Altherr1994}) for the transition rate for a photon of energy
$E$ into an axion of the same energy \citep{Dicus1980,Raffelt1988,Altherr1994}.

White dwarfs were used to constrain the axion-photon coupling (see (\ref{eq-a2-08}) and Fig.~\ref{fig-a2-01}a), and it was noted that the somewhat large period 
decrease of the ZZ~Ceti star G117-B15A, a pulsationally unstable white dwarf, 
could be ascribed to ``invisible'' axion cooling.

On the other hand, under assumptions valid for DAV pulsating white dwarfs 
(comprising ``visible'' cooling and gravitational contraction effects) to which 
G117-B15A belongs, i.e. that neutrino emission is negligible and the star is in 
the evolutionary stage prior to crystallization of its core, the luminosity is

\begin{equation}
L_{*}^{WD} = - \frac{dE_{thermal}}{dt} - \frac{dE_{grav}}{dt} \simeq 6.18 \cdot 10^{30} ~erg/s,
\label{eq-a2-09}
\end{equation}

\noindent
where, as commonly accepted, $E_{thermal}$ denotes the thermal energy of the 
star, and $E_{grav}$ is the fraction of gravitational energy contributing to the
luminosity \citep{Biesiada2004} experimentally measured by \cite{McCook1999}.

An interesting physical problem emerges in certain peripheric layers of such 
white dwarfs: does the possible appearance of the $\gamma$-quanta of axion 
origin, induced by the strong magnetic field in the tachocline via the 
thermomagnetic Ettingshausen-Nernst effect, produce the mixed ``visible'' 
gravothermal (see Eq.(\ref{eq-a2-09})) and ``invisible'' axion (see 
Eq.(\ref{eq-a2-08})) luminosity

\begin{equation}
\log \frac{L_a^{WD} + L_{*}^{WD}}{L_{Sun}} = ?
\label{eq-a2-10}
\end{equation}

In order approach this problem, let us first estimate the strong magnetic field 
in the tachocline.

Subject to a quasi-steady state characterized by a balance of the magnetic field
of the dynamo, in the limit of weak collision, a thermomagnetic current can be 
generated in a (non)degenerate magnetized plasma. For a fully ionized hydrogen 
plasma with $Z = 1$ the  generalized thermomagnetic Ettingshausen-Nernst effect 
leads to a current density (see Eqs.~(\ref{eq06-02})-(\ref{eq06-08})) given by

\begin{equation}
j_y = -\frac{c}{B}\nabla p = - \frac{c}{B} \frac{dp}{dz},
\label{eq-a2-11}
\end{equation}

\noindent
where the pressure $p$ corresponds to the overshoot tachocline (at the base of 
the convective envelope (see Eq.~(\ref{eq06-09}))).

From Maxwell's equation $4\pi j_{\perp}/c = 4\pi j_y /c = curl ~B$, one has

\begin{equation}
j_y = \frac{c}{4\pi} \frac{dB}{dz}
\label{eq-a2-12}
\end{equation}

\noindent
and thus by equating (\ref{eq-a2-11}) and~(\ref{eq-a2-12}) we obtain as a result
of integration in the limits $[B_{OT}, 0]$ in the left-hand side and 
$[0, p_{OT}]$ in the right-hand side:

\begin{equation}
\frac{B_{OT}^2}{8\pi} = p_{OT}
\label{eq-a2-13}
\end{equation}

\noindent
where the pressure $p_{OT}$ in overshoot tachocline (see Eq.~(\ref{eq06-15})) 
is determined by the following cases:

\begin{equation}
p = \begin{cases}
n k_B T, ~nonrelativistic ~ideal ~gas,\\
K_1 \rho ^{5/3}, ~nonrelativistic ~degenerate ~gas,\\
K_2 \rho ^{4/3}, ~relativistic ~degenerate ~gas.
\end{cases}
\label{eq-a2-14}
\end{equation}

From the astronomical data \mbox{\citep{Bhatia2001}} it is known that white dwarfs have
a mass of solar order and planetary sizes, thus the density at the center is 
$10^5 ~g/cm^3 \leqslant \rho_c \leqslant 10^6 ~g/cm^3$, while on the periphery 
it is $10^2 ~g/cm^3 \leqslant \rho_{periph} \leqslant 10^3 ~g/cm^3$. This 
requires considering the quantum behavior of matter. Given that all atoms are 
ionized and so electrons are free, the physical assumption is that it is the 
pressure of this electron gas that balances the gravitational force 
\mbox{\citep{VanHorn1979}}. Such electron gas may be considered degenerate, since the 
temperature, corresponding to Fermi energy $E_F$ is greater than that of the 
white dwarfs\footnote{For example, the Fermi energy $E_F$ of $\sim 1 ~MeV$ 
yields $T \sim 10^{10} ~K$ , which is much greater than the usual internal 
temperature of such stars ($T \sim 10^7 ~K$).}.

Therefore, for a cold magnetic white dwarf the pressure of degenerate 
nonrelativistic electron plasma in (\ref{eq-a2-13}) will be given by

\begin{equation}
p_{OT} = K_1 \rho^{5/3} = \frac{(3 \pi ^2)^{2/3}}{5} \cdot \frac{\hbar ^2}{m_e (\mu _e m')^{5/3}} \cdot \rho^{5/3} \approx 3.12 \cdot 10^{12} \rho^{5/3} ~erg/cm^3
\label{eq-a2-15}
\end{equation}

\noindent
where $m' = \mu _e m_p \approx 2m_p (\mu_e = A/Z \simeq 2)$ and $m_p$ is the 
mass of the proton.

In this regard let us remind that the magnetic field of the thermomagnetic 
current (see Eq.(\ref{eq-a2-11})) in the overshoot tachocline neutralizes the 
magnetic field of the dynamo in core of the stars like ZZ~Ceti G117-B15A.

We use the models of a stably burning hydrogen envelope on a helium core, 
obtained by solving the equations of hydrostatic balance, heat transport, energy
generation, and mass conservation (e.g. \cite{Robinson1995,Steinfadt2010}). 
However, the most important feature between the $H$ and $He$ layers for our 
purposes is not only the chemical profile, but also the neutralization of the 
magnetic fields of the core on the core -- envelope boundary.

Taking the mass density in the tachocline $\rho \leqslant 10^2 ~g/cm^3$ we 
estimate the pressure of a degenerate non-relativistic electron plasma 
(\ref{eq-a2-15}) as

\begin{equation}
\frac{B_{OT}^2}{8\pi} = p_{OT} \leqslant 6.7 \cdot 10^{15} ~erg/cm^3,
\label{eq-a2-16}
\end{equation}

\noindent
where the poloidal magnetic field in tachocline gives rise to value 

\begin{equation}
B_{OT} \leqslant 4.1 \cdot 10^4 ~T = 4.1 \cdot 10^8 ~G.
\label{eq-a2-17}
\end{equation}

Let us consider some properties of the anchored twisted magnetic flux tubes 
depending on the poloidal field in the tachocline through the shear flows 
instability development (see analogous Fig.~\ref{fig-lampochka}a,b). A complete 
cooling of the O-loop inside the twisted magnetic tube near the tachocline leads
to production of the $\gamma$-quanta of axion origin with the probability

\begin{equation}
P_{a \rightarrow \gamma}^{WD} = \frac{1}{4} \left( g_{a \gamma} B_{MS} L_{MS} \right)^2 < 0.022.
\label{eq-a2-18}
\end{equation}

\noindent
where $g_{a\gamma} = 4.4 \cdot 10^{-11} ~GeV^{-1}$ is hadron axion coupling 
constant to photons (see Eq.~\ref{eq-a2-03}), $B_{MS} \leqslant B_z (0) \approx 
B_{OT}$ is the horizontal magnetic field of the O-loop (see 
Fig.~\ref{fig-lampochka}a), $L_{MS} < d$ is the height of the magnetic shear steps, $d \leqslant 
100~km$ is the thickness of the tachocline in magnetic white dwarf 
\citep{Kissin2015}. 

\begin{figure}[tbp]
  \begin{center}
    \includegraphics[width=15cm]{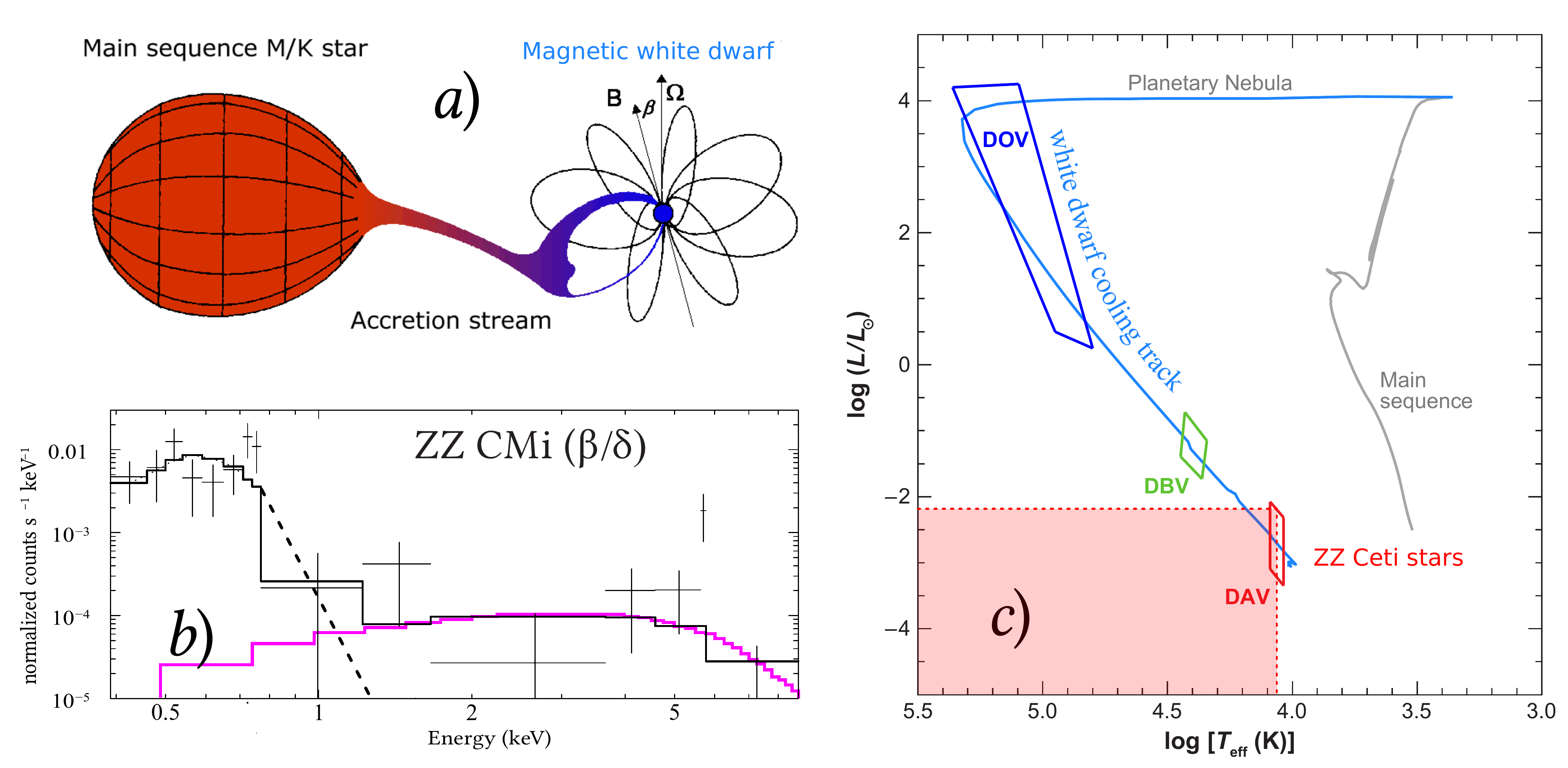}
  \end{center}
\caption{(a) Schematic of binary system. The red dwarf secondary loses gas which
accretes onto the strongly magnetic white dwarf via an accretion stream. At some
point between the two stars the energy density of the magnetic field is 
sufficient to redirect the flow so that accretion takes place near or at the 
magnetic pole. (Adapted from \cite{Cropper1990}).
(b) Swift/XRT spectra of the WD symbiotic with newly discovered X-ray emission 
together with their X-ray spectral types ZZ~CMi. The full line shows the 
best-fit model (black line), while the dotted line shows the contribution of the
individual spectral components in the case of multi-component models. (Adapted 
from \cite{Luna2013}). X-ray emission (rose line) is the $\gamma$-quanta of 
axion origin coming from the magnetic tubes, anchored in tachocline of white 
dwarf ZZ~CMi.
(c) The observed pulsating white dwarf stars lie in three strips in the H-R 
diagram (adopted from \cite{Winget2008}). The estimates of the pulsating DAV 
white dwarf (like ZZ~Ceti star, G117-B15A) parameters are shown here. They enter
the ZZ~Ceti variable (DAV) instability region, a discrete strip in the 
$\log(T_{eff}(K)) - \log(L/L_{Sun})$ plane that spans $T_{eff} \sim 12000~K$ at 
$\log(L/L_{Sun}) = -2.2$ (red dotted line).}
\label{fig-a2-02}
\end{figure}

The theoretically estimated relative portion of the $\gamma$-quanta of axion 
origin is then

\begin{equation}
L_{a \rightarrow \gamma} ^{WD} = P_{a \rightarrow \gamma} ^{WD} L_a ^{WD} < 0.022 \cdot 4.7 \cdot 10^{-3} L_{Sun} \simeq 10^{29} ~erg/s,
\label{eq-a2-19}
\end{equation}

\noindent
which is in agreement with the experimental one (see Fig.\ref{fig-a2-02}b). 
Since it is rather small, the total luminosity is the sum of the ``visible'' 
gravothermal (see Eq.(\ref{eq-a2-09})) and ``invisible'' axion (see 
Eq.(\ref{eq-a2-08}) components

\begin{equation}
\log \frac{L_a ^{WD} + L_* ^{WD}}{L_{Sun}} \simeq -2.2.
\label{eq-a2-20}
\end{equation}

As a result, following our aim, we determined the parameters of hadron 
axion-photon coupling (\ref{eq-a2-03}) and derived several important parameters 
from white dwarf cooling (G117-B15A):

\begin{itemize}
\item \textbf{Luminosity.} Total luminosity~(\ref{eq-a2-20}), depending on the 
axion luminosity through energy-loss rate of a helium plasma at $T = 1.2 \cdot 
10^7 ~K$ of the pulsating white dwarf star G117-B15A by axion emission with 
$g_{a\gamma} = 4.4 \cdot 10^{-11} ~GeV^{-1}$ (Fig.\ref{fig-a2-01}a) and 
gravothermal luminosity, experimentally measured by \cite{McCook1999}, coincides
with the observed pulsating white dwarf stars lying in three strips in the H-R 
diagram (see Fig.~\ref{fig-a2-02}c). Adopted from \cite{Winget2008}). We gave the
estimated parameters of the pulsating DAV white dwarf, or ZZ~Ceti star, 
G117-B15A, which enter the ZZ~Ceti variable (DAV) instability region, a discrete
strip in the $\log(T_{eff}(K)) - \log(L/L_{Sun})$ plane that spans $T_{eff} \sim 
12000~K$ at $\log(L/L_{Sun}) = -2.2$.

\item \textbf{Magnetic field and age.} Strong magnetic field $B_{OT} \sim 
4.1 \cdot 10^8 ~G$, produced in the tachocline by the thermomagnetic 
Ettingshausen-Nernst effect
, and age of G117-B15A, 
when the total luminosity begins to decay (10$^9$~yr) in a $0.55M_{Sun}$ (see 
Fig.\ref{fig-a2-01}b), agree with the theoretical estimates of the magnetic 
field and age of magnetic white dwarfs \citep{Kissin2015}.

\item \textbf{Darkspots by analogy to sunspots.} The spectroscopic observations 
in white dwarfs,  showed the variations of equivalent width in the Balmer lines,
proposing that these variations are due to a star-spot (or darkspot) on the 
magnetic white dwarfs, analogous to a sunspot, which is affecting the 
temperature at the surface, and therefore its photometric magnitude 
\citep{Brinkworth2005,Valyavin2011,Valyavin2014}. According to previous studies,
this variability can be explained by the presence of a dark spot having a 
magnetic nature, analogous to a sunspot. Motivated by this idea, astrophysicists
examine possible physical relationships between the suggested dark spot and the 
strong-field magnetic structure (magnetic ``spot'' or ``tube'') recently 
identified on the surface of this star 
\citep{Maxted2000,Valyavin2011,Valyavin2014}.

It is generally believed \citep{Valyavin2014}, that the magnetic field 
suppresses atmospheric convection, leading to dark spots in the most magnetized 
areas. \cite{Valyavin2014} also found that strong fields are sufficient to 
suppress convection over the entire surface in cool magnetic white dwarfs, which
inhibits their cooling evolution relative to weakly magnetic and non-magnetic 
white dwarfs, making them appear younger than they truly are. Still it should be
kept in mind that this is only a supposition, and not a strict proof, since the 
problem of generation and transport of energy by magnetic flux tubes remains 
unsolved.

In the present paper we show how the problem of energy transport by magnetic 
flux tubes and darkspots formation may be solved by analogy to sunspot by the 
Parker-Biermann cooling effect (see Section~\ref{parker-biermann}). Here we 
imply a theoretical possibility of the time variation of the darkspot activity 
to correlate with the flux of the $\gamma$-quanta of axion origin, which are 
produced within almost horizontal magnetic field of O-loop inside the magnetic 
tube (see Fig.~\ref{fig-lampochka}a).

Because the convection in the convection zone is partial (i.e. it only works 
between the cool region of magnetic tube (see Fig.~\ref{fig-lampochka}a) and 
the surface of magnetic white dwarf), the axion mechanism based on the lossless 
$\gamma$-quanta ``channeling'' along the magnetic tubes allows to explain the 
effect of the almost complete suppression of the convective heat transfer, and 
thus to understand the known puzzling darkness of the starspots (by analogy to 
the sunspots in \cite{Rempel2011}.
\end{itemize}

\section{X-ray coronal luminosity variations}
\label{appendix-luminosity}

\begin{figure*}[htpb!]
\noindent
\centerline{\includegraphics[width=15cm]{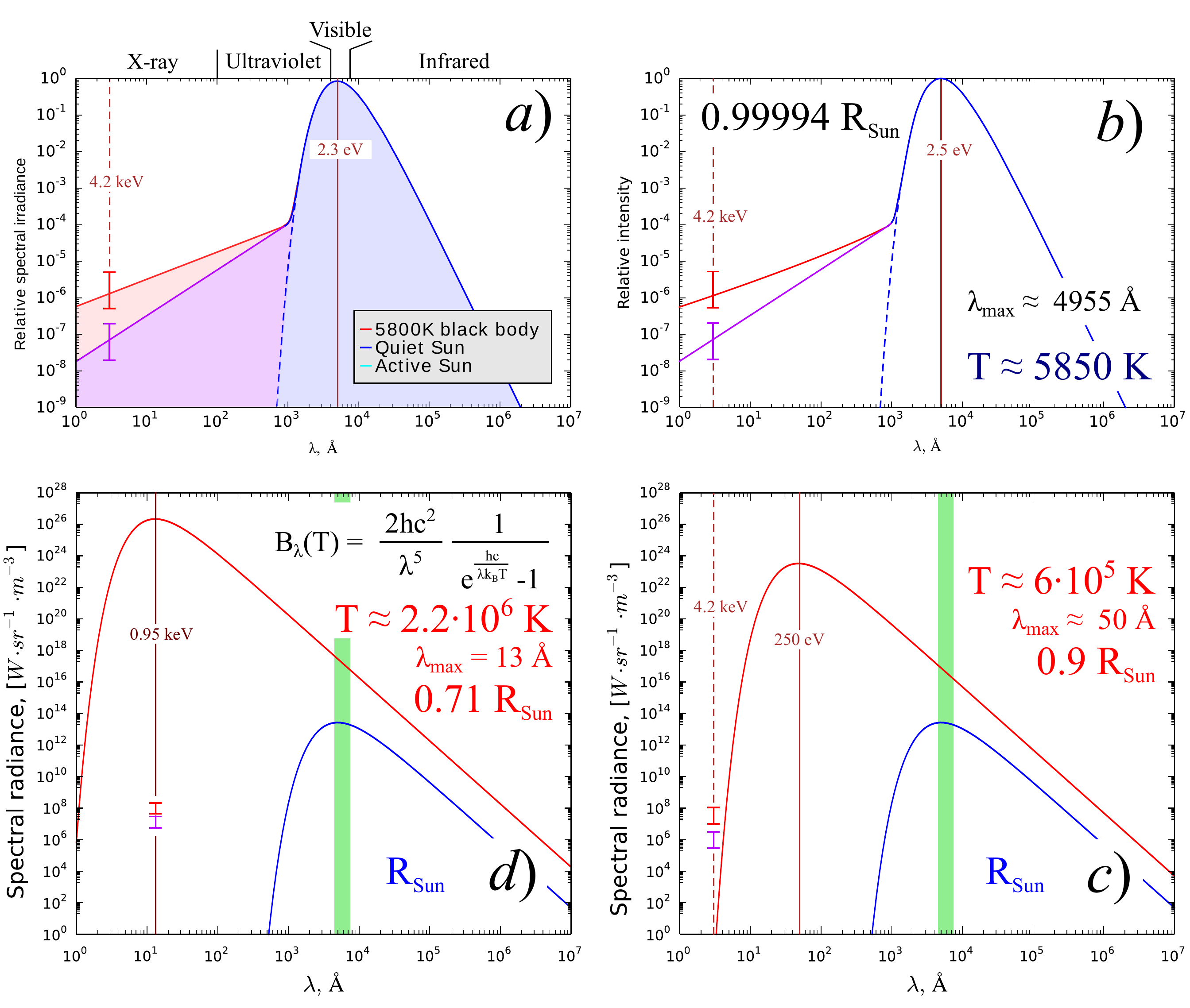}}
\caption{a) The smoothed Solar spectrum corresponds to a black body with a
         temperature $5.77 \cdot 10^3$~K at $R_{Sun}$ (see Fig.~11
         in~\cite{ref36}). 
         The red and violet bars represent
         the high and low contributions from the X-rays of axion origin with mean energy of 
         4.2~keV born in the ``magnetic steps'' at $\sim 0.72 R_{Sun}$,
         roughly estimated for 
         the solar cycle 22. It is possible to perform similar 
         estimations for any other solar cycle using the 
         \mbox{Eqs.~(\ref{eq06-08}), (\ref{eq3.35}), (\ref{app-b-eq03})}
         and the data from \mbox{\cite{Dikpati2008,Pevtsov2014,Lockwood2001}}.
         Note that these X-ray contributions exist within the magnetic tubes 
         only.
         \newline
         b) A spectrum of a black body with a temperature 5850~K at 
         0.99994$R_{Sun}$ (see \mbox{Fig.~\ref{fig-Bz}}).
         The X-ray luminosity (high intensity red and low intensity violet lines) 
         is determined by the Compton scattering of the mentioned 4.2~keV.
         \newline
         c) Red line is a spectrum of a black body with a temperature $6 \cdot 10^5$~K
         at 0.9$R_{Sun}$ (see \mbox{Fig.~\ref{fig-Bz}} and 
         \mbox{\cite{ref45-3}}). 
         The high and low X-ray luminosity (red and violet bars) are determined by the 4.2~keV
          X-rays which propagate along the 
         cool region of the magnetic tube without scattering 
         (\mbox{Fig.~\ref{fig-lampochka}a)}. Blue line corresponds to the black-body spectrum
         of the solar surface.
         \newline
         d) A spectrum of a black body with a temperature
         $2.22 \cdot 10^6$~K at 0.71$R_{Sun}$ (see \mbox{Fig.~\ref{fig-Bz}} and 
         \mbox{\cite{ref45-3}}). The X-ray luminosity
         is determined by the average energy $\sim$0.95~keV thermal photons 
         only (in tachocline \mbox{\citep{Bailey2009}}). These 
         photons are converted into axions in ``magnetic steps'' at 
         $\sim 0.72 R_{Sun}$, and thus do not constitute the spectra of the 
         higher layers. As above, the blue line corresponds to the black-body 
         spectrum of the solar surface.}
\label{app-b-fig01}
\end{figure*}

The X-ray luminosity of the Solar corona during the active phase of the solar
cycle is defined by the following expression:
\begin{align}
\left( L_{corona} ^{X} \right)_{max} = \int \limits_{X-ray} \frac{d\Phi_{corona}^{max} (E)}{dE} \cdot E dE =  \int
\limits_{X-ray} \frac{d W_{corona}^{max} (E)}{dE} dE\, .
\label{app-b-eq01}
\end{align}

In the quiet phase it may be written as
\begin{align}
\left( L_{corona} ^{X} \right)_{min} = \int \limits_{X-ray} \frac{d\Phi_{corona}^{min} (E)}{dE} \cdot E dE =  \int
\limits_{X-ray} \frac{d W_{corona}^{min} (E)}{dE} dE\, .
\label{app-b-eq02}
\end{align}

Then integrating the blue curve in Fig.~\ref{app-b-fig01} for
$\left(L_{corona}^X \right) _{max}$ and the cyan curve for
$\left(L_{corona}^X \right)_{min}$, we obtain

\begin{equation}
\frac{\left( L_{corona}^X \right) _{min}}{L_{Sun}} \sim 10^{-7} ; ~~~~~
\frac{\left( L_{corona}^X \right) _{max}}{L_{Sun}} \sim 10^{-6} .
\label{app-b-eq03}
\end{equation}

So it may be derived from here that the Sun's luminosity is quite low in
X-rays~(\ref{app-b-eq03}), typically (see~\cite{Rieutord2014}),

\begin{equation}
10^{-7} L_{Sun} \leqslant L_{corona}^X \leqslant 10^{-6} L_{Sun},
\label{app-b-eq04}
\end{equation}

\noindent but it varies with the cycle (see blue and cyan lines in
Fig.~\ref{app-b-fig01}a) as nicely shown by the pictures obtained with the
Yohkoh satellite (see Fig.~4 in~\cite{Rieutord2014}).

And finally, it may be supposed that X-rays, propagating from the tachocline
towards the photosphere, interact with the charged particles via the Compton
scattering, but only outside the magnetic tubes. The axion-originated X-ray
radiation channeling inside the magnetic tubes does not experience the Compton
scattering up to the photosphere (Figure~\ref{app-b-fig01}).

\section{Explanation of the high X-ray intensity bands widening near the Yohkoh image edges}
\label{appendix-widening}

It is interesting to note that the bands of high X-ray intensity on Yohkoh
images deviate from the solar parallels (Fig.~\ref{fig-Yohkoh}b). This is
especially the case near the edges of the visible solar disk.

\begin{figure*}
\centerline{\includegraphics[width=15cm]{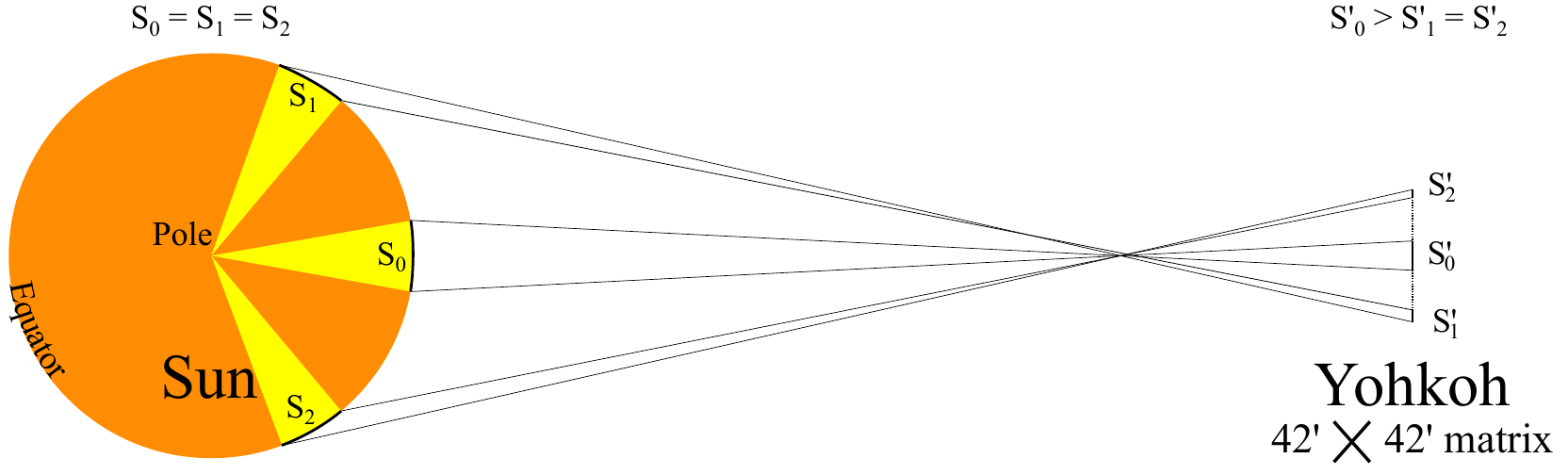}}

\caption{A sketch of the Sun image formation on the Yohkoh matrix. \label{app-a-fig05}}
\end{figure*}

\begin{figure*}[htb!]
\centerline{\includegraphics[width=9cm]{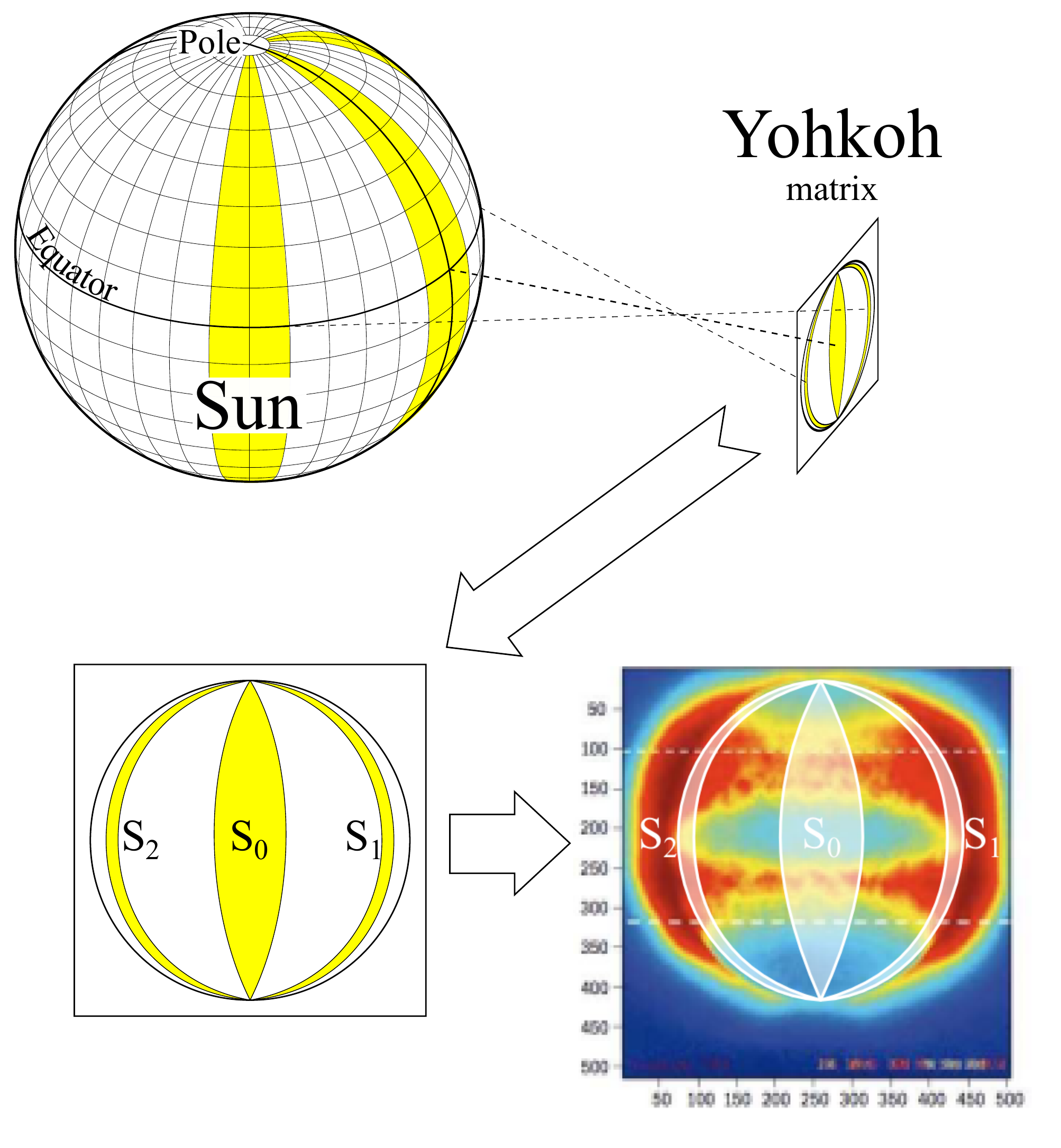}}

\caption{A sketch of the Sun image formation on the Yohkoh matrix. \label{app-a-fig04}}
\end{figure*}

This effect may be explained graphically by means of figures~\ref{app-a-fig05}
and~\ref{app-a-fig04}. These figures show the schematic concept of the Sun
image formation on the Yohkoh matrix. Fig.~\ref{app-a-fig05} shows the Sun from
its pole.

Let us choose three sectors of equal size on the surface of the Sun
(Fig.~\ref{app-a-fig05}, \ref{app-a-fig04}). The areas of the photosphere cut
by these sectors are also equal ($S_0 = S_1 = S_2$). However, as it is easily
seen in the suggested scheme, the \emph{projections} of these sectors on the
Yohkoh matrix are not of equal area (Fig.~\ref{app-a-fig05},
\ref{app-a-fig04}). The $S'_1$ and $S'_2$ projections areas are much less than
that of the $S'_0$ projection (Fig.~\ref{app-a-fig05}). It means that the
radiation emitted by the sectors $S_1$ and $S_2$ of the solar photosphere and
captured by the satellite camera will be concentrated within \emph{less} area
(near the edges of the solar disk) than the radiation coming from the $S_0$
sector (in the center of the solar disk). As a result, the satellite shows
higher intensity near the image edges than that in the center, in spite of the
obvious fact that the real radiation intensity is equal along the parallel of
the Sun.

Therefore, because of the system geometry, the satellite tends to ``amplify''
the intensity near the image edges, and the areas that correspond to the yellow
and green areas at the center (Fig.~\ref{fig-Yohkoh}b) become red near the
edges, thus leading to a visible widening of the high intensity bands.

A particularly high radiation intensity near the very edges of the visible
solar disk, observed even during the quiet phase of the Sun
(Fig.~\ref{fig-Yohkoh}a), indicates a rather ``wide'' directional radiation
pattern of the solar X-rays.

Let us make a simple computational experiment. We will choose a sphere of a
unit radius and spread the points over its surface in such a way that their
density changes smoothly according to some dependence of the \emph{polar} angle
($\theta$). The \emph{azimuth} angle will not influence the density of these
points. For this purpose any function that provides a smooth change of the
density will do. For example, this one:
\begin{equation}
\rho (\theta) = [\rho _0 + \rho _{max} \cdot \cos(2 \cos\theta)]^{-1}\, . \label{app-a-eq01}
\end{equation}

\begin{figure*}[tb!]
\centerline{\includegraphics[width=9cm]{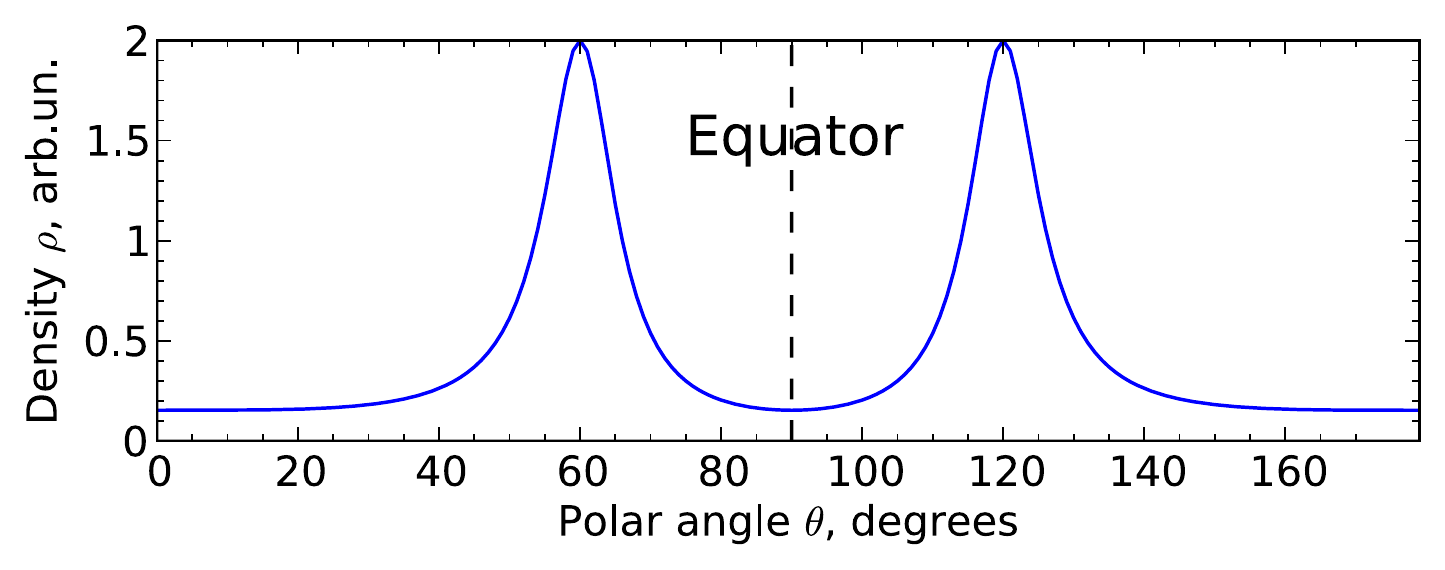}}

\caption{Graphical representation of Eq.~(\ref{app-a-eq01}). \label{app-a-fig06}}
\end{figure*}

Here we take $\rho_0=3.5$ and $\rho_{max}=3$ in arbitrary units. The graphical
representation of this dependence is shown in Fig.~\ref{app-a-fig06}.

I.e. it yields a minimum density of the points near the poles and the equator,
and the maximum density of the points near $\theta = 60^{\circ}$ and
$\theta = 120^{\circ}$.

The polar angle $\theta$ was set in the range $[0, 180^{\circ}]$ with the step
of $1^{\circ}$. The azimuth angle $\varphi$ was set in the range
$[0,180^{\circ}]$ (one hemisphere) with a variable step $\Delta \varphi$
representing the variable density, since the points density is inversely
proportional to the step between them ($\Delta \varphi \sim 1 / \rho$). We
assume that
\begin{equation}
\Delta \varphi (\theta) = \Delta \varphi _0 + \Delta \varphi _{max} \cdot \cos(2 \cos\theta) ~~[deg] .
\label{app-a-eq01a}
\end{equation}

The values of $\Delta \varphi _0 = 3.5^{\circ}$ and
$\Delta \varphi _{max} = 3^{\circ}$ were chosen arbitrarily. So,
\begin{align}
\Delta \varphi (\theta) = 3.5 + 3 \cdot \cos(2 \cos\theta) ~~[deg] .
\label{app-a-eq03}
\end{align}

\begin{figure*}[tb!]
  \begin{center}
  \begin{minipage}[h]{0.49\linewidth}
    \center{a) \includegraphics[width=5cm]{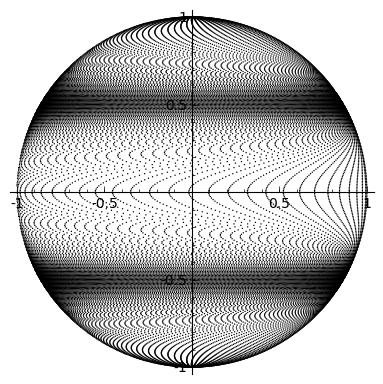}}
  \end{minipage}
  \begin{minipage}[h]{0.49\linewidth}
    \center{b) \includegraphics[width=6cm]{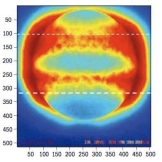}}
  \end{minipage}
  \caption{a) Simulation of the high intensity bands formation on the 2D
           projection of the sphere.
  \newline b) Sun X-ray image from Yohkoh satellite during the active
  phase of the Sun.}
  \label{app-a-fig07}
  \end{center}
\end{figure*}

From Eq.~(\ref{app-a-eq03}) it is clear that the minimum step was 0.5$^{\circ}$
and the maximum step was 6.5$^{\circ}$. Apparently, the more is the step, the
less is the density (near the poles and the equator) and vice versa, the less
is the step, the more is the points density. This was the way of providing a
smooth change of the points density by latitude (along the solar meridians).

Obviously, this forms the ``belts'' of high density of the points along the
parallels. The projection of such sphere on any plane perpendicular to its
equator plane will have the form shown in Fig.~\ref{app-a-fig07}a. As it is
seen in this figure, although the density does not depend on azimuth angle,
there is a high density bands widening near the edges of the \emph{projected}
image. These bands are similar to those observed on the images of the Sun in
Fig.~\ref{app-a-fig07}b.

Let us emphasize that the exact form of the dependence (\ref{app-a-eq01a}) as
well as the exact values of its parameters were chosen absolutely arbitrarily
for the sole purpose of the qualitative effect demonstration. They have no
relation to the actual latitudinal X-ray intensity distribution over the
surface of the Sun.

\bibliographystyle{apj}
\bibliography{Rusov-AxionSunLuminosity}

\end{document}